\documentclass[12pt,preprint]{aastex}

\begin{document}

\title{General Relativistic Magnetohydrodynamic Simulations of Collapsars}
    
\author{Yosuke Mizuno\altaffilmark{1}, Shoichi Yamada\altaffilmark{2}, Shinji Koide\altaffilmark{3} and Kazunari Shibata\altaffilmark{4}}

\altaffiltext{1}{Department of Astronomy, Kyoto University, Sakyo, 
    Kyoto 606-8502, Japan; mizuno@kusastro.kyoto-u.ac.jp}
\altaffiltext{2}{Department of Science and Engineering, Waseda University, Shinjuku, 
    Tokyo 169-8555, Japan; shoichi@heap.phys.waseda.ac.jp}
\altaffiltext{3}{Department of Engineering, Toyama University, Gofuku, 
Toyama 930-8555, Japan; koidesin@ecs.toyama-u.ac.jp}
\altaffiltext{4}{Kwasan and Hida Observatory, Kyoto University, Yamashina, 
    Kyoto 607-8471, Japan; shibata@kwasan.kyoto-u.ac.jp}
    
\shorttitle{GENERAL RELATIVISTIC MHD SIMUATIONS OF THE COLLAPSARS}

\begin{abstract}
We have performed 2.5-dimensional general relativistic magnetohydrodynamic (MHD) simulations of the gravitational collapse of a magnetized rotating massive star as a model of gamma ray bursts (GRBs). 
The current calculation focuses on general relativistic MHD with simplified microphysics (we ignore neutrino cooling, physical equation of state, and photodisintegration). 
Initially, we assume that the core collapse has failed in this star. A few $M_{\odot}$ black hole is inserted by hand into the calculation. The simulations presented in the paper follow the accretion of gas into a black hole that is assumed to have formed before the calculation begins.
The simulation results show the formation of a disk-like structure and the generation of a jetlike outflow inside the shock wave launched at the core bounce. We have found that the jet is accelerated by the magnetic pressure and the centrifugal force and is collimated by the pinching force of the toroidal magnetic field amplified by the rotation and the effect of geometry of the poloidal magnetic field. The maximum velocity of the jet is mildly relativistic ($\sim$0.3c). The velocity of the jet becomes larger as the initial rotational velocity of stellar matter gets faster. 
On the other hand, the dependence on the initial magnetic field strength is a bit more complicated: the velocity of the jet increases with the initial field strength in the weak field regime, then is saturated at some intermediate field strength, and decreases beyond the critical field strength.
 These results are related to the stored magnetic energy determined by the balance between the propagation time of the Alfv\'{e}n wave and the rotation time of the disk (or twisting time). 
 
\end{abstract}

\section{Introduction}

Gamma ray bursts (GRBs) are one of the most enigmatic and most energetic events in the universe (e.g., Klebesadel, Strong, \& Olson 1973; Fishman \& Meegan 1995; van Paradijs, Kouveliotou, \& Wijers 2000). GRBs and the afterglows are well described by the fireball model (e.g., Piran 1999; M\'{e}sz\'{a}ros 2002), in which a relativistic outflow is generated from a compact central engine. Rapid temporal decay of several afterglows is consistent with the evolution of a highly relativistic jet with bulk Lorentz factors $\sim 10^{2} - 10^{3}$ (e.g., Sari, Piran, \& Halphen 1999). The formation of relativistic jets from a compact central engine remains one of the major unsolved problems in GRB models. 

What is the central engine of GRBs? 
From recent observations, some evidence was found for a connection between GRBs and the death of massive stars. 
Analyses of host galaxies show a correlation between star-forming regions and the position of GRBs inside the host galaxy (Bloom, Kulkarni, \& Djorgovski 2002). A \lq\lq bump" resembling that seen in light curves of Type Ic supernovae has also been detected in the optical afterglow of several GRBs (e.g., Bloom et al. 2002; Garnavich et al. 2003). GRB 980425 has been associated with an optical supernova, SN 1998bw (e.g., Iwamoto et al. 1998; Woosley, Eastman, \& Schmidt 1999). From the early-phase observation of a very bright afterglow (e.g., Uemura et al 2003; Price et al. 2003) it has been revealed that GRB 030329 is likely to be in association with SN 2003dh (Stanek et al. 2003; Hjorth et al. 2003). Several authors (Frail et al. 2001; Panaitescu \& Kumar 2001; Bloom, Frail, \& Kulkarni 2003) have studied beaming angles and energies of a number of GRBs. They have found that central engines of GRBs release supernova-like energies ($\sim 10^{51}$ erg). It is thus probable that a major subclass of GRBs is a consequence of the collapse of a massive star. 

One of the most attractive scenarios involving massive stars is the collapsar model (Woosley 1993; MacFadyen \& Woosley 1999). A \lq\lq collapsar" is a rotating massive star. The collapsar model is divided into two classes according to the formation history of the black hole. A \lq\lq type I collapsar" is a failed supernova. The collapse of an iron core leads temporarily to  a neutron star formation. A black hole, however, is formed quickly within a few seconds as a result of the accretion of matter through the stalled shock wave. In the mean time, infalling envelope matter is slowed by rotation in the equatorial plane and forms a disk. 
 The formation and propagation of relativistic flows from a type I collapsar have been studied numerically in both Newtonian (MacFadyen \& Woosley 1999) and special relativistic hydrodynamic simulations (Aloy et al. 2000; Zhang, Woosley, \& MacFadyen 2003). The other model is a \lq\lq type II collapsar," in which a black hole is formed over a longer period of time accompanying a successful supernova. The supernova generates an outgoing shock and ejects all the helium and heavy elements outside the neutron star. However, some of the post-shock gas fails to reach escape velocity and is pulled back toward the proto-neutron star by gravity. When enough gas has fallen back, the neutron star collapses to a black hole. If the infalling matter has sufficient angular momentum, it forms a disk at the same time. The formation and propagation of relativistic flows from a type II collapsar have been studied numerically in Newtonian hydrodynamic simulations (MacFadyen, Woosley, \& Heger 2001). 
However, these previous numerical simulations of collapsar models do not fully address the outflow mechanism. 
In these simulations, in fact, the authors estimate the energy of a jet assuming neutrino annihilation or magnetohydrodynamic (MHD) process as a source and input a jet with that energy from the inner boundary.
There have been a lot of studies of astrophysical jets with Newtonian (e.g., Uchida \& Shibata 1985; Shibata \& Uchida 1986, 1990; Kudoh, Matsumoto, \& Shibata 1998; Kato, Kudoh, \& Shibata 2002) and relativistic MHD simulations (e.g., Koide, Shibata, \& Kudoh 1998, 1999). They have fully addressed the formation, acceleration, and collimation of jets from accretion disks. We think that the formation of relativistic flow from collapsars should also be studied by MHD simulations. In fact, several authors (Cameron 2001; Wheeler, Meier, \& Wilson 2002) proposed that the relativistic jets are generated by the MHD process when the massive star collapses to a rapidly rotating neutron star.

It is suspected that large-scale magnetic fields play an important role in the formation of a GRB. Magnetic fields are suitable for extracting energy on the burst time-scale from the debris disk in the collapsar scenario. If the field is initially weak, it is likely to be exponentially amplified as a result of differential rotation and/or dynamo. Large-scale magnetic fields can help guide and collimate the outflow and also contribute to its acceleration (e.g., Usov 1994; Thompson 1994; Meszaros \& Rees 1997; Katz 1997; Klu\'{z}niak \& Ruderman 1998; Lyutikov \& Blackman 2001; Drenkhahn \& Spruit 2002; Vlahakis \& K\"{o}nigl 2001, 2003ab).  Even if the flow is not magnetically driven, the field should be strong enough to account for the observed synchrotron emission (Spruit, Daigne \& Drenkhahn 2001). If the outflow is largely Poynting flux dominated, it can solve the baryon contamination problem. Magnetic fields have been favored in this respect for driving GRB outflows. 

The effect of stellar rotation and intrinsic magnetic fields on gravitational collapse of massive stars was numerically studied by several authors (LeBlanc \& Wilson 1970; Symbalisty 1984; Ardeljan et al. 2000). Symbalisty (1984) computed the magnetorotational core collapse of a $15 M_{\odot}$ star by numerically solving the 2.5-dimensional MHD equations together with the neutrino transport. The simulations showed the formation of two oppositely directed, high-density, supersonic jets in the combination of a rapid rotation and a very strong dipole magnetic field. Khokhlov et al. (2001) assumed that the jets are generated by a magnetorotational mechanism when a stellar core collapses into a neutron star and simulated the process of the jet propagation through the star. Proga et al. (2003) simulated the collapsar model by using a pseudo-Newtonian MHD code including some essential microphysics (physical equation of state [EOS], photodisintegration of nuclei, and neutrino cooling) for GRBs.

In this study we perform 2.5-dimensional general relativistic MHD(GRMHD) simulations of the gravitational collapse of a rotating star with magnetic field as a model for a collapsar. The collapsar is in some sense an anisotropic supernova, and it is considered that relativistic jets from collapsars are launched by MHD processes in accreting matter and/or by neutrino annihilation. We investigate the physics of the formation of jets, the acceleration force on jets, and the dependence of the acceleration of jets on the initial magnetic field strengths and on the initial rotational velocity. We describe the numerical method in our simulations in section 2 and present our results in section 3. The summary and discussion are given in section 4.

\section{Numerical Method}

\subsection{Basic Equations}

In order to study the formation of relativistic jets from a collapsar, we use a 2.5-dimensional GRMHD code with spherical polar coordinates ($r, \theta, \phi$) (i.e., $\partial /\partial \phi = 0$, but $v_{\phi}$ and $B_{\phi}$ are nonzero; Koide 2003. The method is based on the general relativistic formulation of the conservation laws of particle-number and energy-momentum, Maxwell equations, and Ohm's law with no electrical resistance (ideal MHD condition) on a curved space-time (Weinberg 1972; Koide et al. 1998, 1999; Koide 2003). 

The space-time $(x^{0}, x^{1}, x^{2}, x^{3})=(ct,x^{1},x^{2},x^{3})$ is described by the metric $g_{\mu \nu}$, where the line element $ds$ is given by $(ds)^{2}= g_{\mu \nu} dx^{\mu}dx^{\nu}$. Here $c$ is the speed of light. The basic equations of GRMHD in four-dimensional space-time are
\begin{equation}
\nabla_{\nu}(\rho U^{\nu}) = {1 \over \sqrt{-||g||}}{\partial \over \partial x^{\nu}}(\sqrt{-||g||} \rho U^{\nu}) =0, \label{eq1}
\end{equation}
\begin{equation}
\nabla_{\nu} T^{\nu \mu} ={1 \over \sqrt{-||g||}}{\partial \over \partial x^{\nu}}(\sqrt{-||g||} T^{\mu \nu}) + \Gamma^{\mu}_{\sigma \nu} T^{\sigma \nu} =0,\label{eq2}
\end{equation}
\begin{equation}
\partial_{\mu} F_{\nu \lambda} + \partial_{\nu} F_{\lambda \mu} +\partial_{\lambda} F_{\mu \nu} =0,\label{eq3}
\end{equation}
\begin{equation}
\nabla_{\mu} F^{\mu \nu} = -\mu_{0} J^{\nu},\label{eq4}
\end{equation}
where $\nabla_{\nu}$ is the covariant derivative, $||g||$ is the determinant of $g_{\mu \nu}$ (as a matrix), and $\Gamma^{\lambda}_{\mu \nu} \equiv 
{1 \over 2}g^{\lambda \sigma} \left( -{\partial g_{\mu \nu} \over \partial x^{\sigma}} + {\partial g_{\nu \sigma} \over \partial x^{\mu}} +{\partial g_{\sigma \mu} \over \partial x^{\nu}} \right)$ are the Christoffel symbols. $U^{\nu}$ and $J^{\nu} = (c \rho_{e}, J^{1}, J^{2}, J^{3})$ are four-velocity and four-current density, respectively ($\rho_{e}$ is the electric charge density). The general relativistic energy momentum tensor $T^{\mu \nu}$ is given as
\begin{equation}
T^{\mu \nu} = pg^{\mu \nu} +(e_{int} + p) U^{\mu} U^{\nu} +F^{\mu}_{\sigma} F^{\nu \sigma} -{1 \over 4} g^{\mu \nu} F^{\lambda \kappa} F_{\lambda \kappa}
\end{equation}
where $F^{\mu \nu}$ is the electromagnetic field-strength tensor, $F_{\mu \nu} =\partial_{\mu} A_{\nu} -\partial_{\nu} A_{\mu}$ and $A^{\mu} =(\Phi_{e}/c, A^{1}, A^{2}, A^{3})$ is the four-vector potential ($\Phi_{e}$ is the electro-static potential). The electric field $E_{i}$ and the magnetic field $B_{i}$ are given by $E_{i} = c F_{i 0}$ ($i = 1, 2, 3$) and $B_{1} = F_{23}$, $B_{2} = F_{31}$, $B_{3} = F_{12}$, respectively. The electric field $E_{i}$ and the magnetic field $B_{i}$ are normalized as $E_{i} = E_{i}^{\star}/\sqrt{\mu}$ and $B_{i} = B_{i}^{\star}/\sqrt{\mu}$, respectively, where $\mu$ is the magnetic permeability and quantities with asterisks are given in MKSA units (SI units). The scalars $\rho$, $p$, and $e_{int}$ are proper mass density, proper pressure, and proper internal energy density, respectively. We assume mainly for numerical simplicity that matter can be described as an ideal gas; hence, $e_{int}= \rho c^{2} + p/(\Gamma -1)$, where $\Gamma$ is the specific heat ratio although we know that the gamma law EOS is not very appropriate for the gravitational collapse of massive stars. We further assume that $\Gamma =5/3$ in the simulations for numerical reasons although it is more appropriate to adopt $\Gamma = 4/3$ for the gas of current interest.
In addition to the equations, we assume the infinite electric conductivity condition
\begin{equation}
F_{\mu \nu} U^{\nu} = 0.\label{eq6} 
\end{equation}
Then, equations (\ref{eq1})-(\ref{eq3}) together with equation (\ref{eq6}) are a closed system of equations. Equation (\ref{eq4}) is used only to calculate the current density $J^{\mu}$. 

If we assume that the off-diagonal elements of the metric $g_{\mu \nu}$ vanish, 
\begin{equation}
g_{\mu \nu} = 0 \quad (\mu \neq \nu),
\end{equation}
and we use the notation
\begin{equation}
g_{00} = -h_{0}^{2}, \quad g_{11}=h_{1}^{2}, \quad g_{22}=h_{2}^{2}, \quad g_{33} = h_{3}^{2}, 
\end{equation}
then the line element can be written as
\begin{equation}
(ds)^{2} = g_{\mu \nu} dx^{\mu} dx^{\nu} = -h^{2}_{0} (cdt)^{2} + \sum^{3}_{i=1}h^{2}_{i}(dx^{i})^{2}.
\end{equation}

We employ the 3+1 formalism (Thorne, Price, \& Macdonald 1986; Koide et al. 1999; Koide 2003). The velocity of light is written explicitly. In the non-relativistic limit ($c \to \infty $), they reduce to the standard Newtonian MHD equations. 

In the GRMHD code, a simplified total variation diminishing (TVD) method is employed (Davis 1984). This method is useful because it requires only the maximum speed of physical waves but not the eigenvalues of the Jacobian of the equations.

De Villiers \& Hawley (2003) and Gammie, McKinney, \& T\'{o}th (2003) have developed a new GRMHD code independently. De Villiers \& Hawley (2003) used a non-conservative scheme that is based on techniques developed by Hawley, Smarr, \& Wilson (1984) for axisymmetric hydrodynamics around black holes. On the other hand, Gammie, McKinney, \& T\'{o}th (2003) used a conservative, shock-capturing scheme. They have performed some one- and two- dimensional tests and have obtained consistent results with each other. 

We do not consider the evolution of the metric because the accreted mass is sufficiently small in the time-scale of the simulations. 
The current calculation focuses on GRMHD with simplified microphysics. Although photodisintegration of bound nuclei and cooling by neutrino emission (thermal and especially Urca processes) have been shown to be of critical importance for the dynamics of collapsar accretion disks (Popham, Woosley \& Fryer 1998; MacFadyen \& Woosley 1999), we ignore these effects in the current calculations for simplicity. It is hoped that future GRMHD simulations will include essential microphysics (neutrino cooling, physical EOS, photodisintegration) to realistically simulate collapsar accretion disks.

\subsection{Initial Condition}

As for the initial model, we have the collapsar model in mind. In principle, we should start calculations from a realistic progenitor model with rotation and magnetic field. Although some progress has been made lately on the precollapse evolution of massive stars with rotation and magnetic field (Heger et al. 2003), their models have a lot of uncertainty yet. Hence, we take the following pragmatic approach: 
Core collapse is assumed to fail in this star, and a few $M_{\odot}$ black hole is inserted by hand into the calculation. The simulations presented in the paper follow the accretion of gas into a black hole assumed to have formed before the calculation begins. Furthermore, We use Bruenn's realistic one-dimensional supernova model(Bruenn 1992) as a rough guide to what the density structure exterior to the black hole might be. The gas in Bruenn's model must have collapsed before the present calculation began to form the black hole that is assumed to be present. However, we use the scaling in that model as a guide for the scale-free stellar structure. We employ only the profiles of the density, pressure, and radial velocity as our initial condition. In this way, we can discuss generic features of the dynamics.

\begin{figure}
\plotone{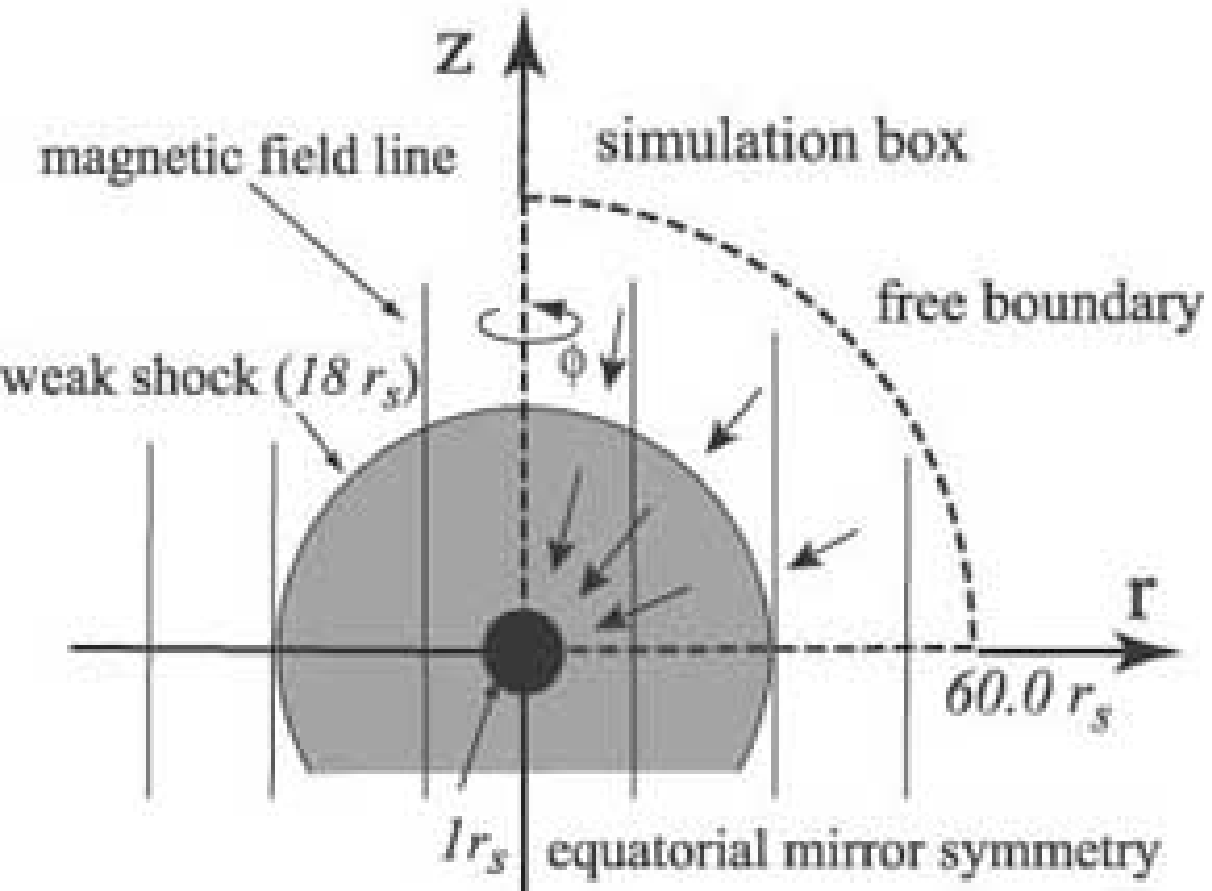}
\caption{Schematic picture of our simulations. Dashed lines show the simulation box. The gray region shows post-shock matter. The black region shows the black hole. \label{fig1}}
\end{figure}

\begin{figure}
\plotone{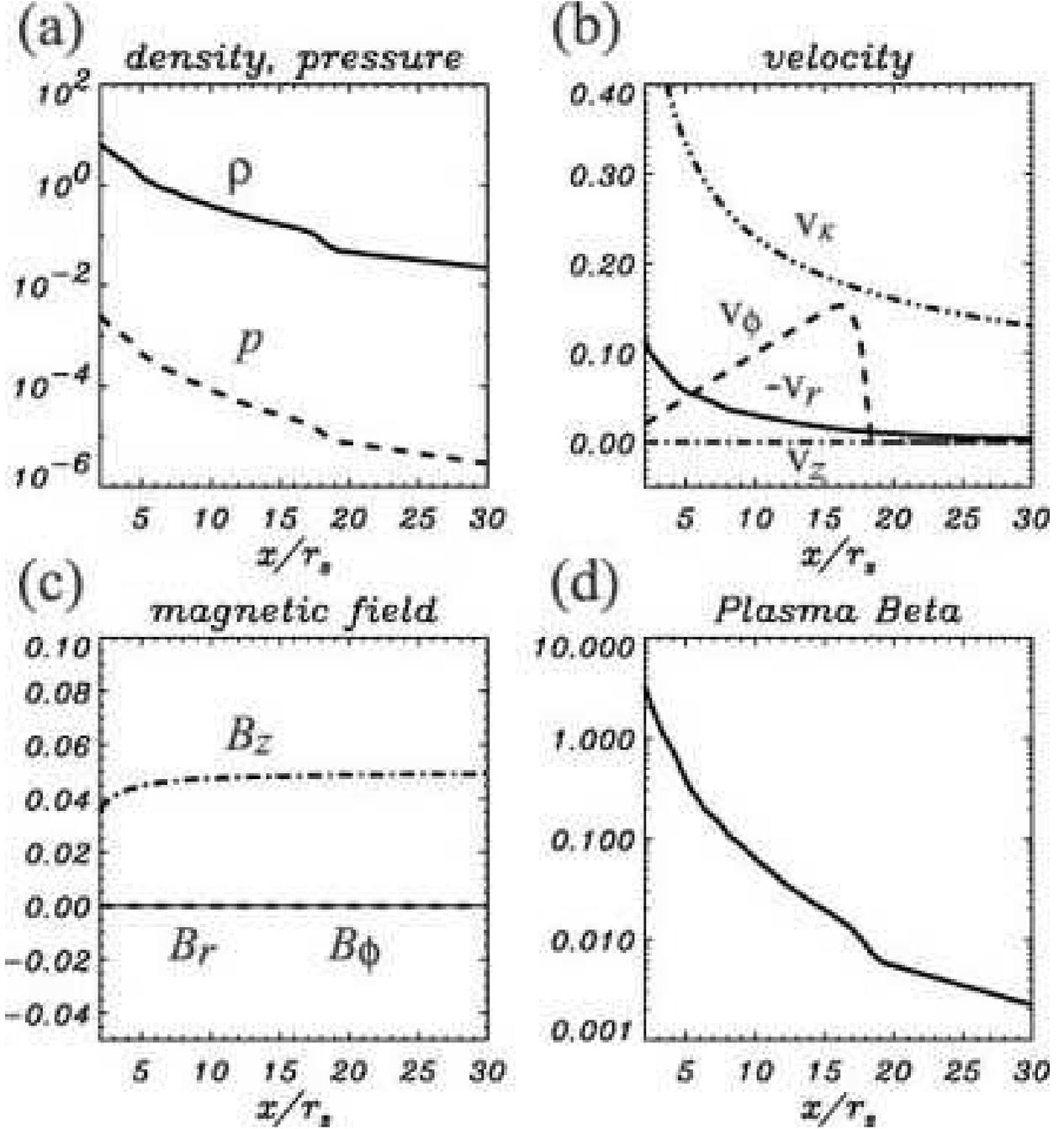}
\caption{Various physical quantities on the equatorial plane $z = 0$ at the initial time in the standard case A2 ($B_{0} = 0.05, v_{0} = 0.01$). {\it (a)} Density $\rho$ {\it (solid line)} and gas pressure $P$ {\it (dashed line)}. {\it (b)} Components of velocity $v_{r}$ {\it (solid line)}, $v_{\phi}$ {\it (dashed line)}, and $v_{z}$ {\it (dot-dashed line)}; $v_{K}$ {\it (double-dot-dashed line)} is the Keplerian velocity. {\it (c)} Components of magnetic field $B_{r}$ {\it (solid line)}, $B_{\phi}$ {\it (dashed line)}, and $B_{z}$ {\it (dot-dashed line)}. (d) Plasma beta $ \beta \equiv P_{gas} / P_{mag}$. The stellar matter falls toward the black hole and rotates like a rigid body around the black hole. \label{fig2}}
\end{figure}

We consider a non-rotating black hole. The metric is given by
\begin{equation}
g_{00}=1-{r_{S} \over R},\quad  g_{11}=\left( 1 -{r_{S} \over R} \right)^{-1},\quad g_{22}=R^{2},\quad g_{33}=R^{2} \sin \theta,
\end{equation}
where $r_{S} \equiv 2 G M_{\odot} / c^{2}$ is the Schwarzschild radius. 
We add the effect of stellar rotation and intrinsic magnetic field in the simulations as the initial condition. The initial rotational velocity distribution is assumed to be a function of the distance from the rotation axis, 
\begin{equation}
r = R \sin \theta,
\end{equation}
 only:
\begin{equation}
v_{\phi} = v_{0} {x_{0}^{2} \over r^{2} + x_{0}^{2}} r \label{eq11} .
\end{equation}
Here, $v_{0}$ is a model parameter for rotational velocity. We fix $x_{0} = 100 \ r_{S}$ in this paper. 
The initial magnetic field is assumed to be uniform and parallel to the rotational axis. This is known as the Wald solution for a non-rotating black hole: $B_{R}=B_{0} \cos \theta, B_{\theta}=- \alpha B_{0} \sin \theta $. $B_{0}$ is the overall magnetic field strength and $\alpha = g_{00}$ is the lapse function.  
Figure \ref{fig1} shows the schematic picture of our simulation. 
The distributions of various physical quantities (density $\rho$, pressure $p$, each velocity component $v_{r}, v_{\phi}, v_{z}$, each magnetic field component $B_{r}, B_{\phi}, B_{z}$, and plasma beta $\beta = p_{gas} / p_{mag}$) on the equatorial plane for the initial state are shown in Figure \ref{fig2}. In the distribution of density and pressure, we can see that the weak shock is sitting at about 18 $r_{S}$. We input the stellar rotation up to 18 $r_{S}$ because the accreted mass for stellar matter in the pre-shock region is sufficiently small in the time scale. We neglect rotation in the pre-shock region for numerical simplicity.
The rotation is almost uniform (see Eq. [\ref{eq11}]). This initial rotation profile is similar to the previous simulations of rotational core collapse (M\"{o}nchmeyer \& M\"{u}ller 1989; Yamada \& Sato 1994).

We want to emphasize that our simulations are scale free. The normalization units are summarized in Table \ref{table1}. We use typical values for normalizations. If we set a black hole mass $ M_{BH} \approx 3 M_{\odot}$, the Schwarzschild radius $r_{S}$ is about $8 \times 10^{5} $ cm and the time unit $\tau_{S} (\equiv r_{S} / c)$ is $3.0 \times 10^{-5} $ s.
The units of magnetic field strength and pressure depend on the normalization of the density. If we take, for example, the density unit $\rho_{0} = 10^{10} {\rm g/cm^{3}}$, the magnetic field strength unit is $3.0 \times 10^{14}$ G and the pressure unit is $ P = 10^{31} {\rm dyn/cm^{2}}$. 
The models computed in this paper are summarized in Table \ref{table2}. These units are the values at the last stable Keplerian orbit of a non-rotating black hole on the equatorial plane [$(r, z) = (r_{0}, 0)$]. The variables with subscript \lq\lq 0" represent the values at the last stable orbit throughout this paper. 

We also use the following nondimensional parameters:
\begin{equation}
E_{th}= {V_{s0}^{2} \over V_{K0}^{2}},
\end{equation}
\begin{equation}
E_{mag}= {V_{A0}^{2} \over V_{K0}^{2}},
\end{equation}
\begin{equation}
E_{rot}= {V_{\phi}^{2} \over V_{K0}^{2}},
\end{equation}
where $V_{s0}$ is the relativistic sound speed,
\begin{equation}
V_{s0}= c \sqrt{{\Gamma p_{0} \over \rho_{0} c^{2} + \Gamma p_{0} / (\Gamma -1)}},
\end{equation}
$V_{A0}$ is the Alfv\'{e}n velocity,
\begin{equation}
V_{A0} = c {B_{0} \over \sqrt{\rho_{0} c^{2} + \Gamma p_{0} /(\Gamma -1) + B_{0}^{2}}},
\end{equation}
and $V_{K0}$ is Keplerian velocity,
\begin{equation}
V_{K0} = c /[2(r_{0}/r_{S}-1)]^{1/2}.
\end{equation}
They are all evaluated at the last stable orbit. 
The values of the parameters adopted in this paper are summarized in Table \ref{table2}.

The simulations are done in the region $2 r_{S} \le R \le 60 r_{S}$, $0 \le \theta \le \pi/2$, with $120 \times 120$ mesh points. We assume axisymmetry with respect to $z$-axis and mirror symmetry with respect to the equatorial plane. We employ a free boundary condition, which waves, fluids and magnetic fields can pass through freely, at $R=2 r_{S}$ and $R=60 r_{S}$ (see Koide et al. 1999).

\section{Numerical Results}

\subsection{Formation of Jet}

\begin{figure}
\plotone{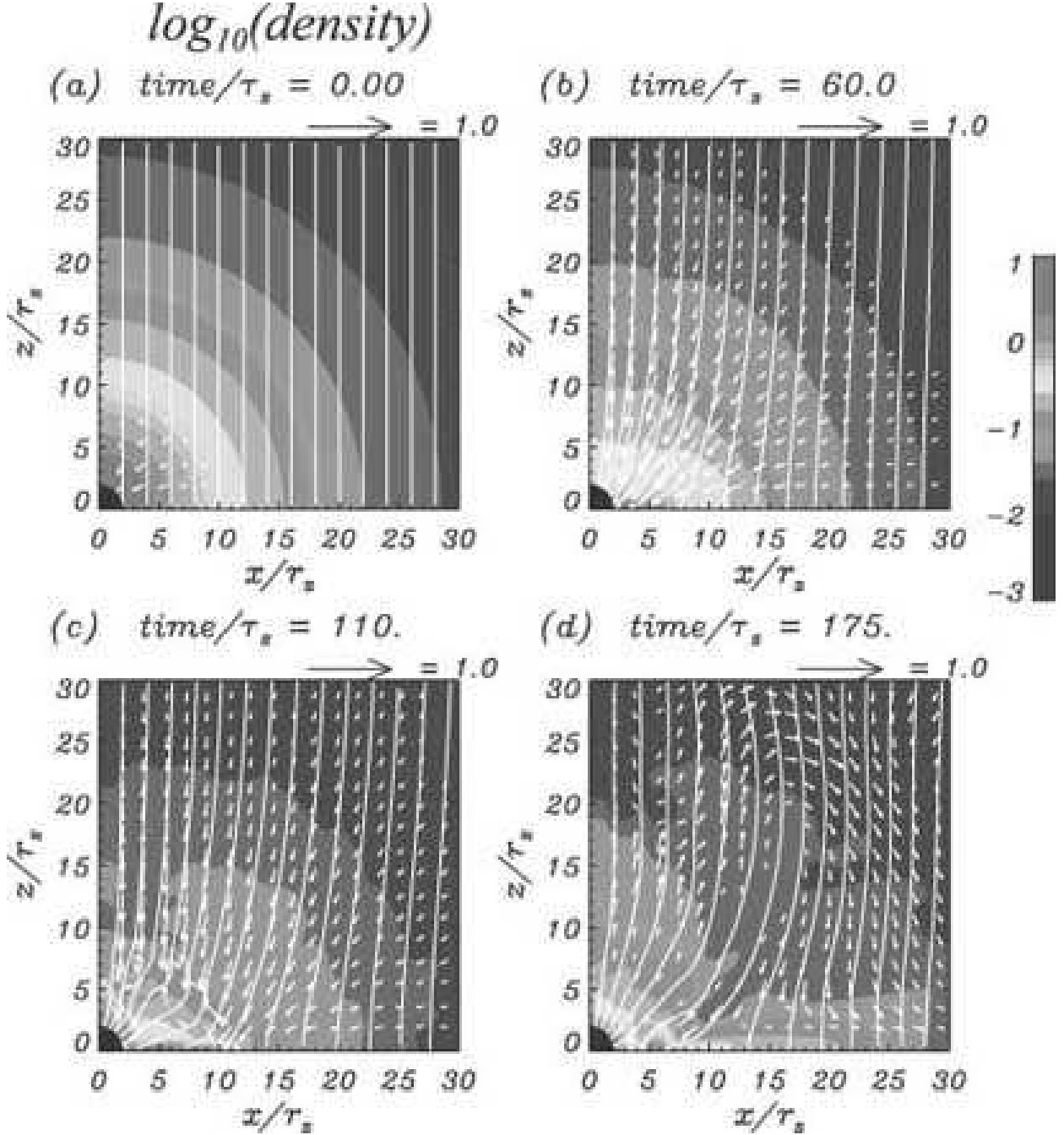}
\caption{Time evolution of the density for case A2 ($B_{0} = 0.05, v_{0} = 0.01$). The color scale shows the value of the logarithm of density. The white curves depict magnetic field lines. Arrows represent the poloidal velocities normalized by the light velocity. The central black region corresponds to the inner boundary at $r/r_{S} = 2$. The distance and the time are given in units of $r_{S}$ and $\tau_{S} \equiv r_{S}/c$, respectively. {\it (a)} Initial condition. The stellar matter falls and rotates around the black hole. {\it (b)} Condition at $t/\tau_{S} = 60$. Accreting matter forms a disklike structure. {\it (c)} Condition at $t/\tau_{S} = 110$. The shock wave is generated near the black hole. {\it (d)} Condition at $t/\tau_{S} = 175$. The shock wave propagates outward. The jetlike outflow is produced inside the shock wave. 
\label{fig3}}
\end{figure}

We shall first discuss the results of the case A3 (Table \ref{table2}; from now on referred to as the standard case). The model parameters are $B_{0} = 0.05$, $v_{0} = 0.01$, $E_{th}  = 2.36 \times 10^{-3}$, $E_{mag} = 1.68 \times 10^{-3}$, and $E_{rot} = 5.36 \times 10^{-2}$. Figure \ref{fig3} shows the time evolutions of density. The stellar matter falls onto the central black hole at first. This collapse is anisotropic as a result of the effects of rotation and magnetic field. The accreting matter  falls more slowly on the equatorial plane than on the rotational axis. The matter piles up on the equatorial plane, and a disklike structure is formed near the central black hole. At $t/\tau_{S} = 120$, the disk density is about 2 orders of magnitude higher than the density of matter above the disk. Since the magnetic field is frozen into the plasma, it is dragged by the accreting matter and deformed to be more radial. At about $t/\tau_{S} = 60$, a shock wave is formed near the central black hole and propagates outward anisotropically, because of the geometry of the magnetic field. A vortex pattern is generated inside the shock front, and moves outwardly with the propagating shock front, and produces a jetlike outflow. As this jetlike outflow moves up along the magnetic field lines, it becomes collimated. The simulation is terminated at $t/\tau_{S} = 175$, at which time the jet is still propagating outward.

\begin{figure}
\plotone{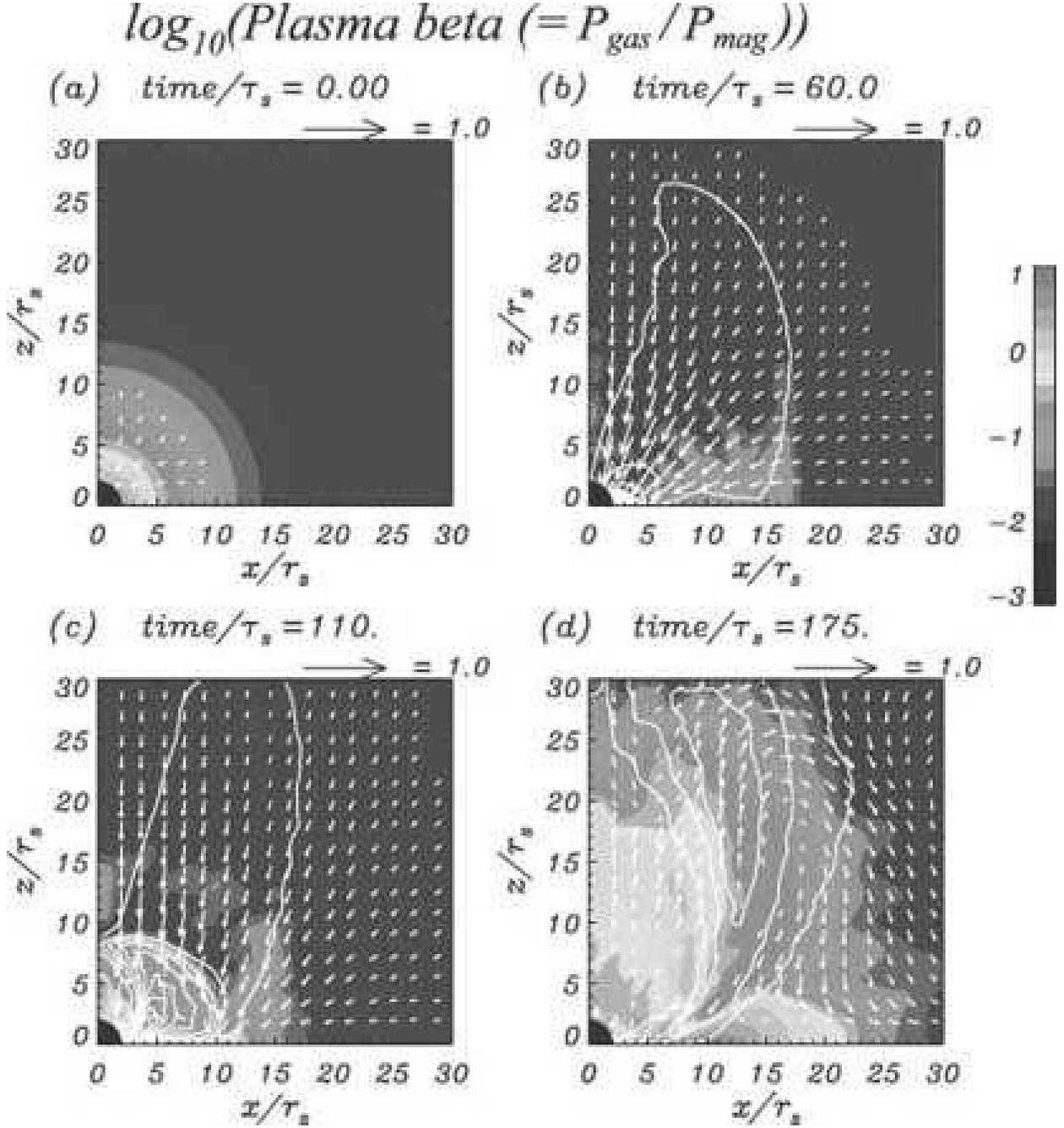}
\caption{The time evolution of the plasma beta ($\beta = P_{gas} / P_{mag}$) distribution for the standard case A2. The color scale shows the value of the logarithm of plasma beta. The white contour plots are the toroidal component of the magnetic field. Arrows show the poloidal velocities normalized by the light velocity. The central black region corresponds to the inner boundary. The distance and the time are given in units of $r_{S}$ and $\tau_{S} \equiv r_{S}/c$, respectively. The amplified magnetic field expands outward and collimates the jetlike outflow. \label{fig4}}
\end{figure}

The jetlike outflow is generated behind the shock wave. Figure \ref{fig4} shows the time evolutions of plasma beta ($= P_{gas}/P_{mag}$) and toroidal magnetic field for the standard case A3. The shock wave expands with twisted magnetic fields, and the outflow has high magnetic pressures. The jetlike outflow formed in the simulation is thus magnetically driven.

The magnetic field plays an important role also in the collimation. 
 The plasma beta is initially low in the whole region, indicating that the magnetic pressure is dominant over the gas pressure (see Fig.(\ref{fig4})). Not only the pinching force of the toroidal magnetic field but also the geometry of the poloidal magnetic field (i.e. poloidal magnetic pressure) plays a crucial role. 

\subsection{Mechanism of Shock}

\begin{figure}
\epsscale{0.9}
\plotone{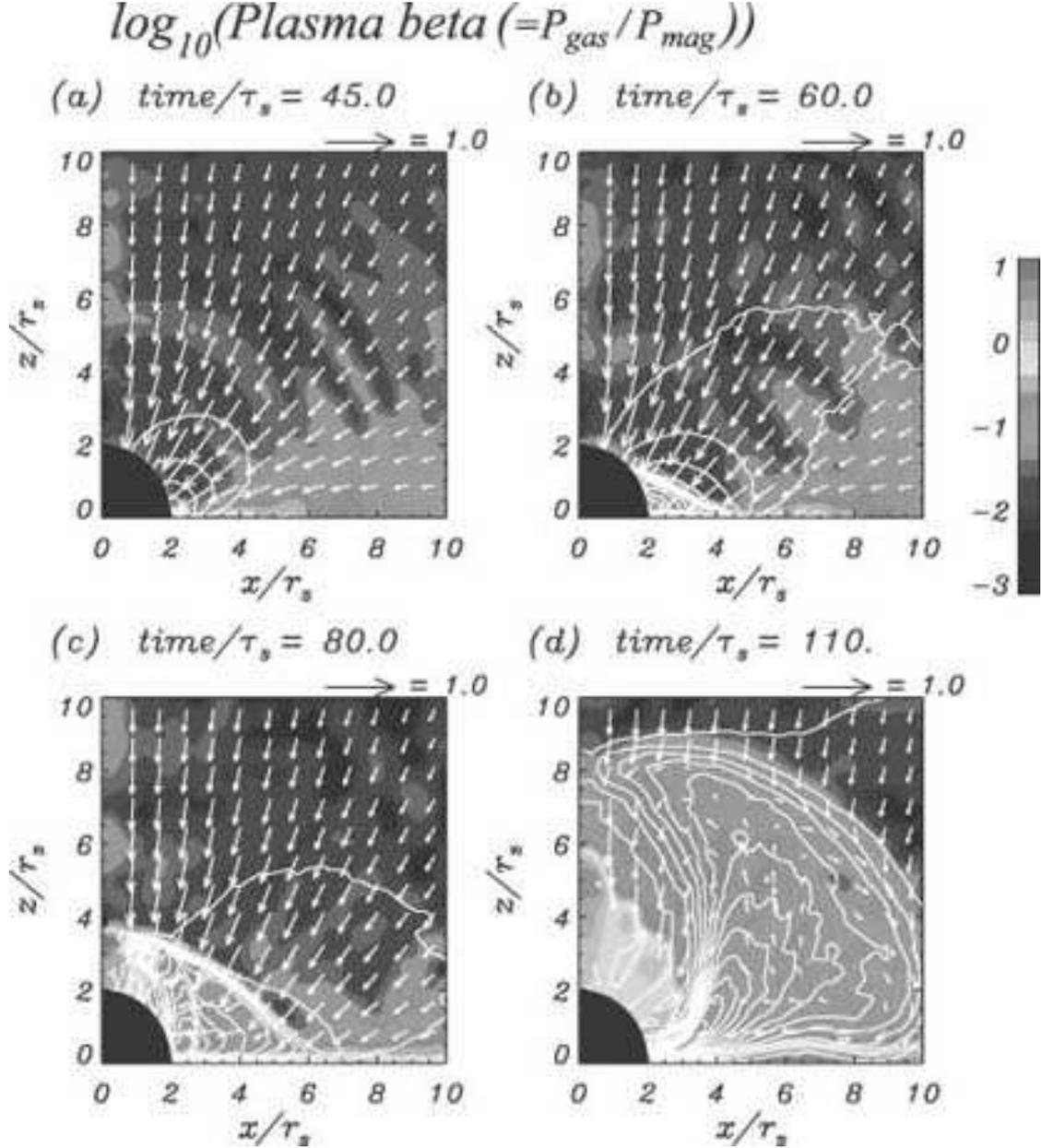}
\caption{Time evolution of the plasma beta ($\beta = P_{gas}/P_{mag}$) distribution near the central black hole for case A2. The color scale indicates the value of the logarithm of plasma beta. The white contour plot shows the toroidal component of the magnetic field. Arrows present the poloidal velocities normalized by the light velocity. The distance and the time are given in units of $r_{S}$ and $\tau_{S} \equiv r_{S}/c$, respectively. {\it (a)} Condition at $t/\tau_{S} = 45$. The toroidal magnetic field is amplified near the black hole. The plasma beta is low in the whole region. {\it (b)} Condition at $t/\tau_{S} = 60$. The shock wave is generated near the black hole. {\it (c)} Condition at $t/\tau_{S} = 80$. The shock wave propagates outwards with amplified toroidal magnetic fields. {\it (d)} Condition at $t/\tau_{S} = 110$. The jetlike outflow is generated inside the shock wave. The plasma beta is low in the post-shock region. \label{fig5}}
\end{figure}

We now investigate in more detail the mechanism of shock formation in the simulation. 
Figure \ref{fig5} shows the time evolution of the spatial distributions of plasma beta ($\beta = P_{gas}/ P_{mag}$) and toroidal magnetic field near the black hole in the standard case A2. At $t/\tau_{S} = 60.0$, the accreting stellar matter rotates faster than earlier around the central black hole. 
As time goes on, the magnetic field is more and more twisted as a result of the differential rotation of accreting matter after the poloidal magnetic field is deformed to be more radial by the radial accretion. The toroidal magnetic field is amplified significantly near the central black hole. The amplified magnetic field expands outward as Alfv\'{e}n waves and launches an outgoing shock wave. 
We can see that the expansion of the toroidal magnetic field takes place at the same position as the shock front. The plasma beta ($\beta = P_{gas} / P_{mag}$) inside the shock wave is low ($\beta < 1$), implying that the shock wave is generated mainly by the magnetic field. 

\begin{figure}
\plotone{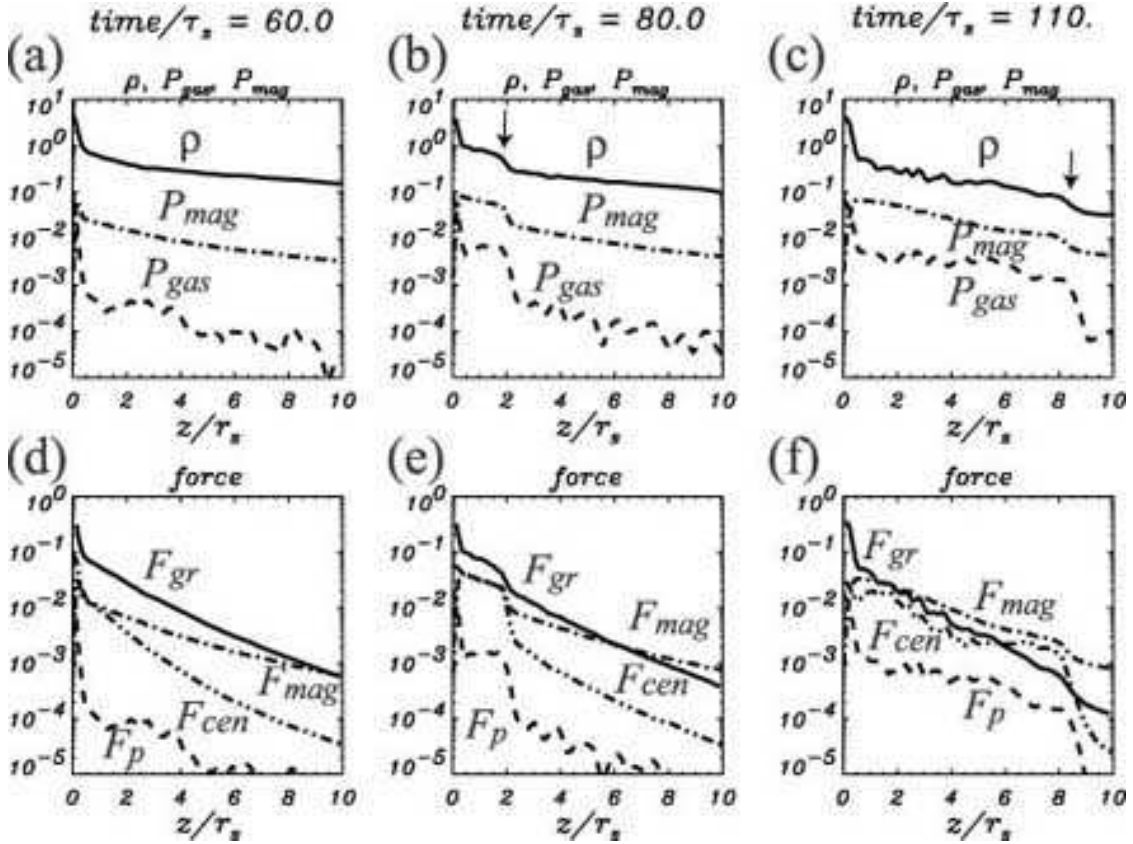}
\caption{Time evolutions of the various physical quantities on the cylindrical surface of $x = 4.0 r_{S}$. ({\it a, b, c}) Density $\rho$ ({\it solid line}), gas pressure $P_{gas}$ ({\it dashed line}), and  magnetic pressure $P_{mag}$ ({\it dot-dashed line}). {\it (d, e, f)} Gravitational force $F_{gr}$ ({\it solid line}), gas pressure gradient force $F_{p}$ ({\it dashed line}), magnetic force $F_{mag}$ ({\it dot-dashed line}), and centrifugal force $F_{cen}$ ({\it double-dot-dashed line}). The times are the same as in Figure \ref{fig5}. These panels show the propagation of the shock (see arrows). In the post-shock region, density, gas pressure, and magnetic pressure increase. In {\it (f)}, the magnetic and centrifugal forces are larger than the gravitational force.  
  \label{fig6}}
\end{figure}

\begin{figure}
\plotone{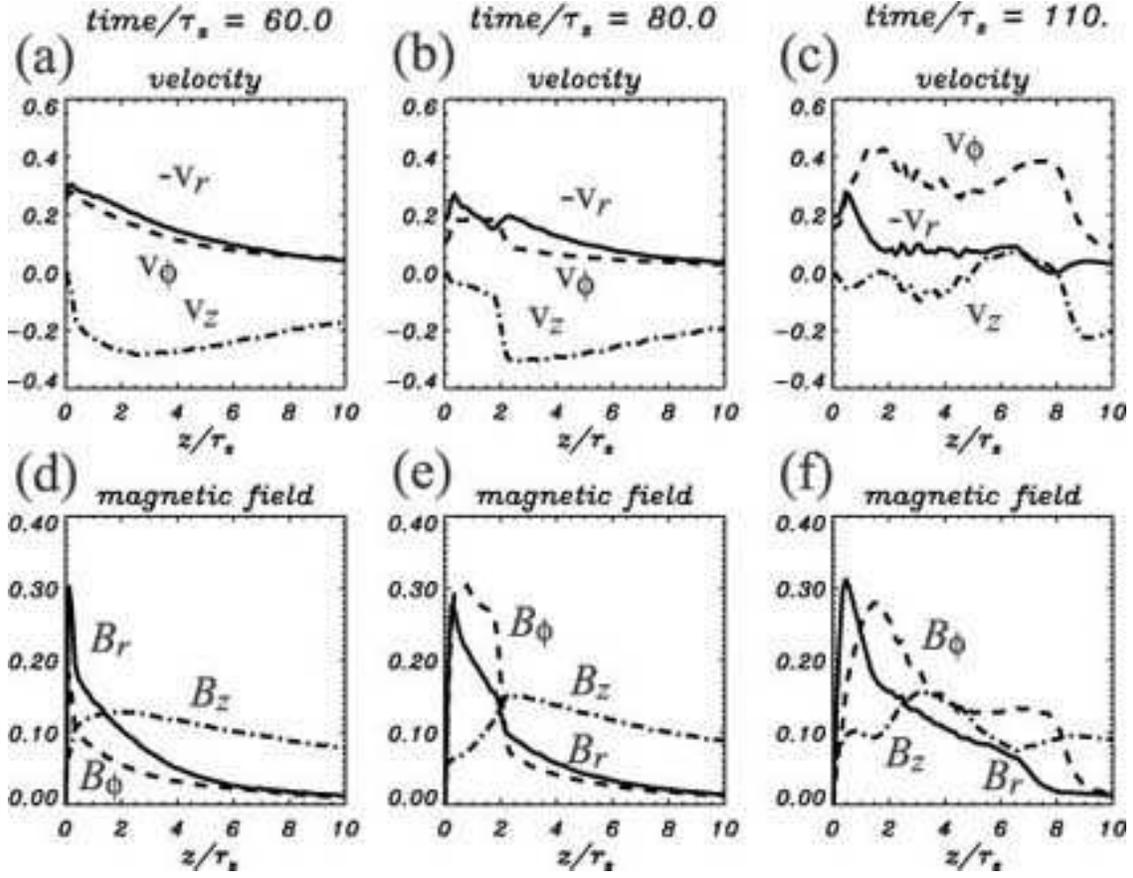}
\caption{Time evolution of the various physical quantities on the cylindrical surface of $x = 4.0 r_{S}$. {\it (a, b, c)} show the components of velocity, $v_{r}$ {\it (solid line)}, $v_{\phi}$ {\it (dashed line)}, and $v_{z}$ {\it (dot-dashed line)}. {\it (d, e, f)} Components of magnetic field, $B_{r}$ {\it (solid line)}, $B_{\phi}$ {\it (dashed line)}, and $B_{z}$ {\it (dot-dashed line)}. The times are the same as in Figure \ref{fig5} and \ref{fig6}. The amplified magnetic field expands outward. Accreting matter is pushed back and the jetlike outflow is generated. \label{fig7}}
\end{figure}

Figure \ref{fig6} and \ref{fig7} show the time evolutions of various physical quantities (density, gas and magnetic pressures, magnetic field, velocity, and forces) on the cylindrical surface with $x/r_{S} = 4$.
It shows the propagation of the outgoing shock. It is also found that when the shock wave is generated near the central black hole, the magnetic field is amplified by about a factor of 6 from the initial strength of the magnetic field near the black hole. 
The magnetic force $F_{mag}$ is comparable to the centrifugal force $F_{cen}$. These are about an order of magnitude higher than the pressure gradient force $F_{p}$. This confirms the claim that the magnetic force, together with the magnetocentrifugal force produces the shock wave rather than the pressure gradient force. 

\subsection{Properties of Jet}

\begin{figure}
\plotone{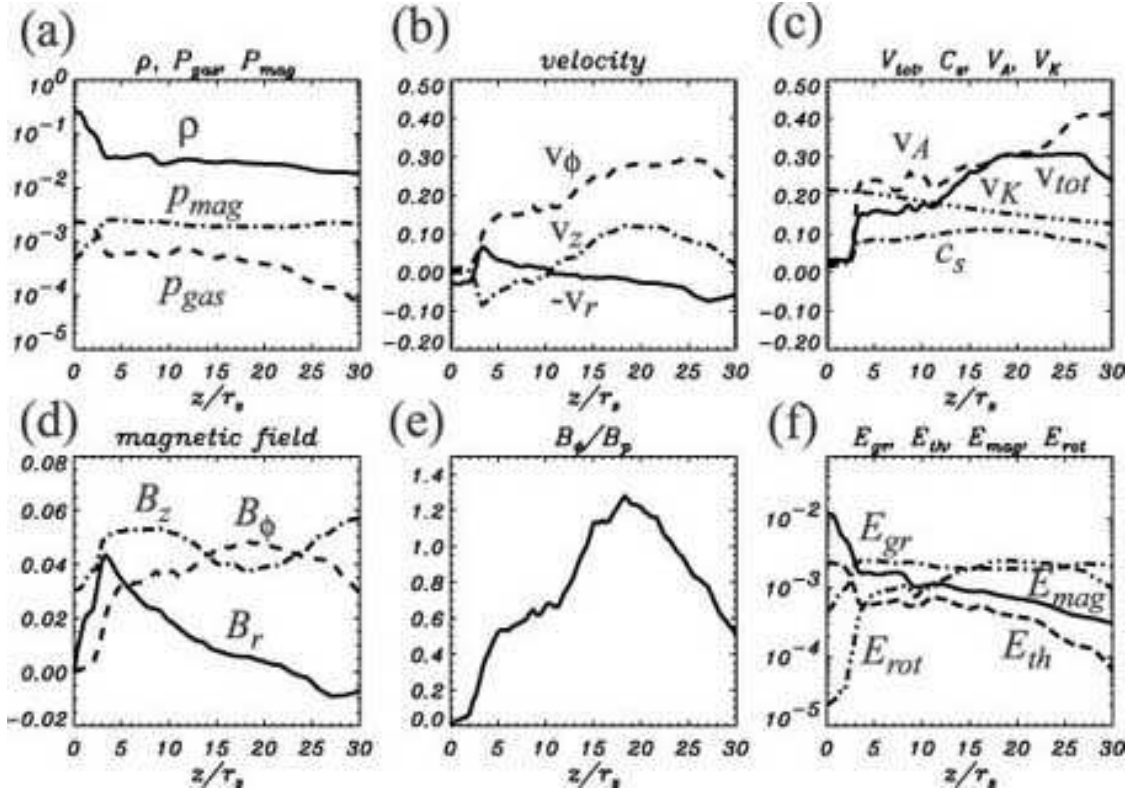}
\caption{Various physical quantities on the surface of $x/r_{S} = 12$ at $t/\tau_{S} = 175$ for case A2. {\it (a)} Density $\rho$ {\it (solid line)}, gas pressure $P_{gas}$ {\it (dashed line)}, and magnetic pressure $P_{mag}$ {\it (dot-dashed line)}. {\it (b)} Components of velocity, $v_{r}$ {\it (solid line)}, $v_{\phi}$ (dashed line), and $v_{z}$ {\it (dot-dashed line)}. (c) Matter velocity $v_{tot}$ {\it (solid line)}, Alfv\'{e}n velocity $v_{A}$ {\it (dashed line)}, sound velocity $c_{S}$ {\it (dot-dashed line)}, and Keplerian velocity $v_{K}$ {\it (double-dot-dashed line)}. {\it (d)} Components of magnetic field, $B_{r}$ {\it (solid line)}, $B_{\phi}$ {\it (dashed line)}, and $B_{z}$ {\it (dot-dashed line)}. {\it (e)} Ratio of toroidal to poloidal magnetic field components $ B_{\phi} / B_{p}$. {\it (f)} Gravitational energy $E_{gr}$ {\it (solid line)}, thermal energy $E_{th}$ {\it (dashed line)}, magnetic energy $E_{mag}$ {\it (dot-dashed line)}, and rotational energy $E_{rot}$ {\it (double-dot-dashed line)}. The jetlike outflow has a helical structure. The maximum jet velocity is mildly-relativistic ($\sim 0.3$ c). However, it is clearly larger than the escape velocity. \label{fig8}}
\end{figure}

We discuss here the properties of the jet found in our simulation. 
Figure \ref{fig8} shows the distributions of various physical quantities along the jetlike outflow at $t/\tau_{S} = 175$. By this time, the shock wave has already passed through the upper boundary ($z/r_{S} = 30$) of this figure. 

It is easily seen in the velocity distribution that the jet has a mildly relativistic velocity, $ \sim 0.3$ c (the poloidal velocity is $ \sim 0.1$ c). It is larger than the sound velocity, as well as the Keplerian velocity. It also clearly exceeds the escape velocity at that point, and the jet is likely to get out of the stellar remnant. The toroidal velocity is larger than the poloidal velocity in the jet-forming region. 
The density of the jet is about twice as high as the density of the surrounding matter, $\sim 3 \times 10^{8} \ \mathrm{g/cm^{3}}$ if the density unit is $\rho_{0} = 10^{10} \ \mathrm{g/cm^{3}}$. 
The toroidal magnetic field is the dominant component of the magnetic field in the jet-forming region. The total magnetic field strength there is $\sim 3.0 \times 10^{13} \ \mathrm{G}$. 

On the other hand, we see in the energy distribution that the magnetic energy is comparable with the rotational energy and larger than the gravitational energy, which is consistent with the previous statement that the jet is accelerated by the magnetic and centrifugal forces. 

\begin{figure}
\epsscale{0.75}
\plotone{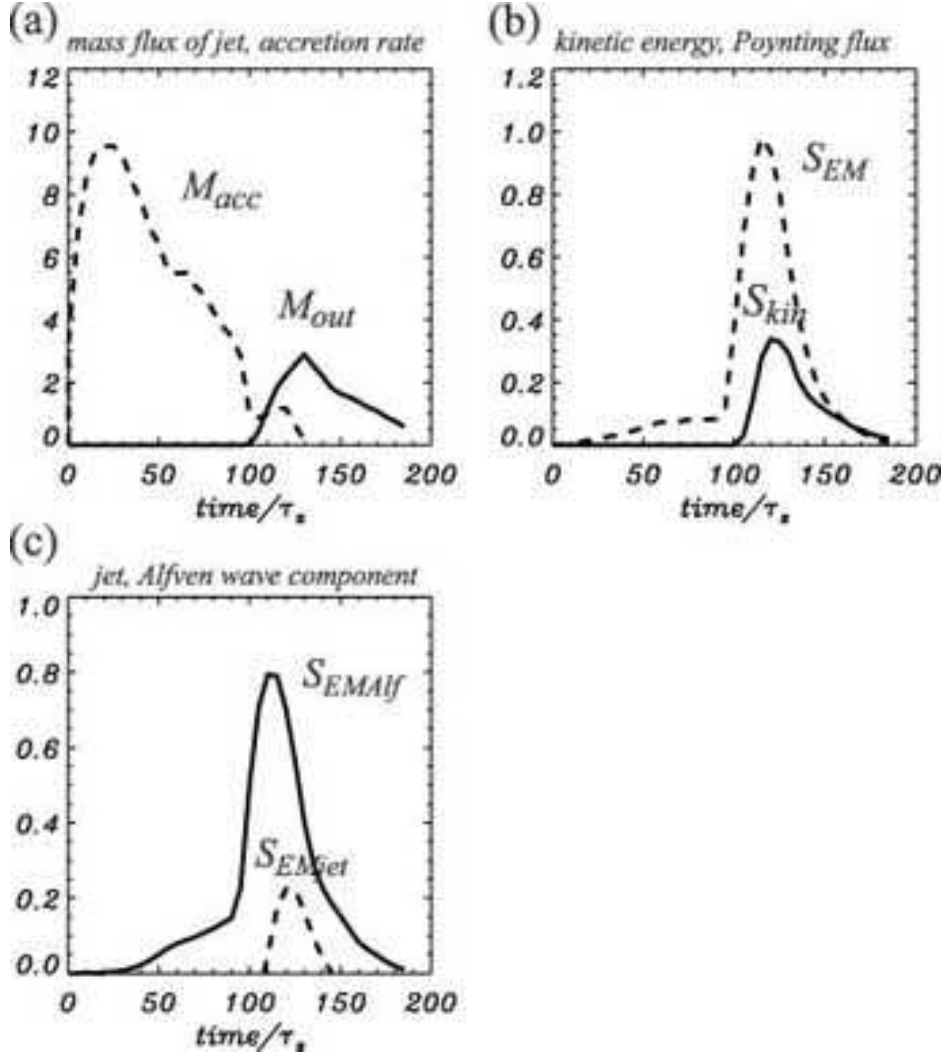}
\caption{Time variation of  {\it (a)} the mass flux of the jet {\it (solid line)} and the accretion rate {\it (dashed line)}, {\it (b)} the kinetic energy {\it (solid line)} and the Poynting flux {\it (dashed line)}, and {\it (c)} the jet {\it (solid line)} and Alfv\'{e}n wave components {\it (dashed line)} of the Poynting flux at $z/r_{S} \simeq 12$ for the standard case A2. \label{fig9}}
\end{figure}

Figure \ref{fig9} shows the time variation of the mass flux of the jet and the accretion rate, the kinetic energy and the Poynting flux, and the jet and Alfv\'{e}n wave components of the Poynting flux at $z/r_{S} \simeq 12$ for the standard case A2. Here, the jet and Alfv\'{e}n components are defined as 
\begin{eqnarray}
S_{EMjet} &=& {B_{\phi}^{2} \over 4 \pi} v_{z} , \\
S_{EMAlf} &=& {B_{\phi} B_{z} \over 4 \pi} v_{\phi} .
\end{eqnarray}
The Poynting flux is an order of magnitude larger than the kinetic energy flux. In the Poynting flux, the Alfv\'{e}n wave component dominates over the jet components. This means that the Alfv\'{e}n wave transports more energy outward. Using the simulation data, we can estimate that the kinetic energy of the jet, $E_{jet}$, is $ \sim 10^{49} \ \mathrm{ergs}$ if the density of the jet is $ 10^{10} {\rm g/cm^{3}}$. This is 2 orders of magnitude lower than the standard energy of GRBs ($\sim 10^{51} \ \mathrm{ergs}$). If Poynting flux is considered, the total energy becomes $\sim 10^{50} \ \mathrm{ergs}$, still less than the energy of GRBs.

\subsection{Dependence on the Initial Magnetic Field Strength}

\begin{figure}
\epsscale{0.75}
\plotone{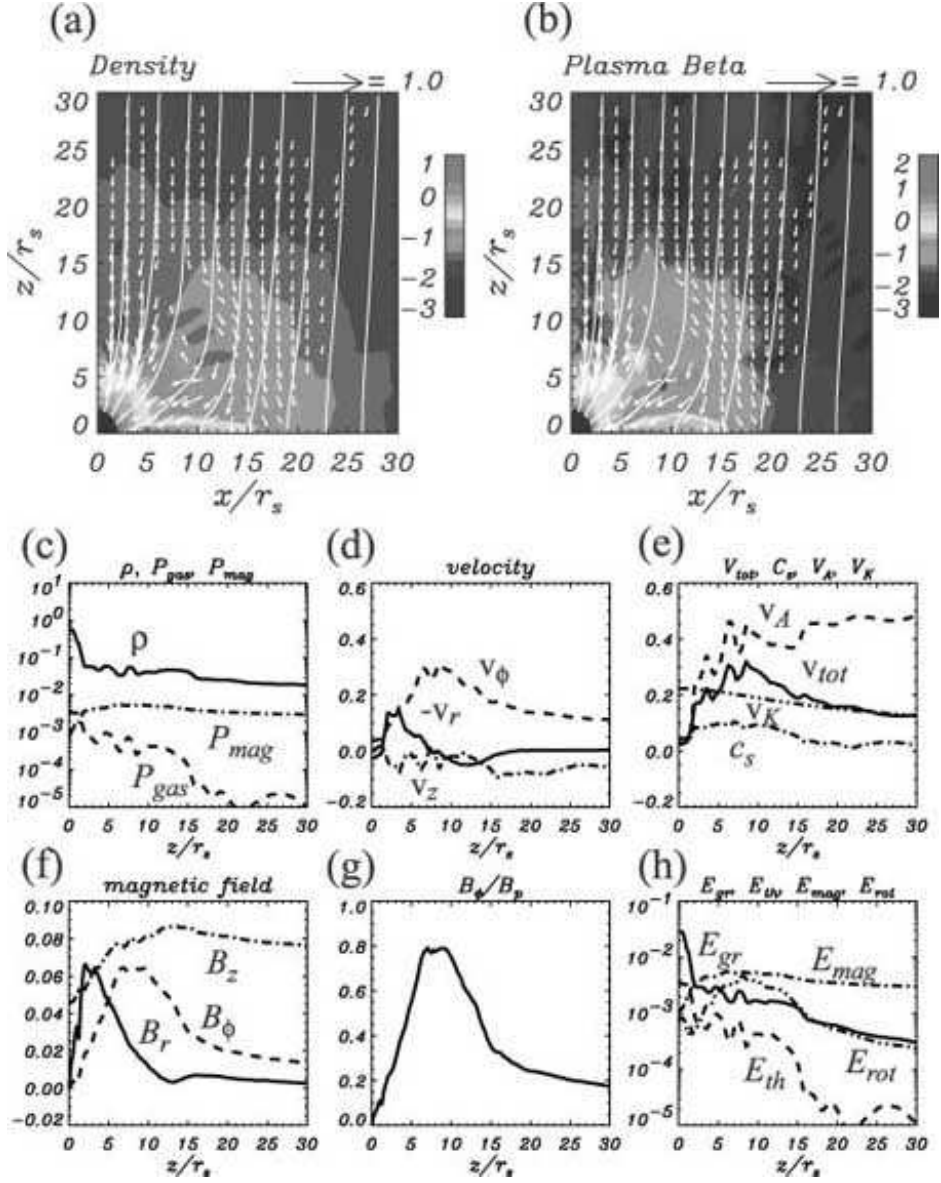}
\caption{Strong magnetic field case A3 ($B_{0} = 0.07$, $v_{0} = 0.01$). 
{\it (a, b)} Snapshots of density and plasma beta at $t/\tau_{S} = 120$. The color scales show the values of the logarithm of density and plasma beta. The white curves represent the magnetic field lines. Arrows depict the poloidal velocities normalized by the light velocity. The central black region corresponds to the inner boundary. The distance and the time are given in units of $r_{S}$ and $\tau_{S} \equiv r_{S}/c$, respectively. {\it (c, d, e, f, g, h)} Various physical quantities on the surface of $x/r_{S} = 11$ at $t/\tau_{S} = 120$. The jetlike outflow is weaker and fainter (the maximum poloidal velocity of the jet is $< 0.05$ c ) than in the standard case A2. The magnetic twist is also weaker. \label{fig10}}
\end{figure}

The dependence of the jet properties on the initial magnetic field strength has also been investigated.
Figure \ref{fig10} shows the density and plasma beta distributions for case A3 in which the initial magnetic field is greater than the standard case A2 ($B_{0} = 0.07, v_{0} = 0.01$).
The jetlike outflow is weaker and fainter than in the standard case A2. The maximum poloidal velocity of the jet is less than 0.05 c, although the jet velocity is still larger than the escape velocity. Thus, even in this case, a jet will emerge. The shock wave itself, as well as the magnetic twist, is also weaker. 

\begin{figure}
\epsscale{0.85}
\plotone{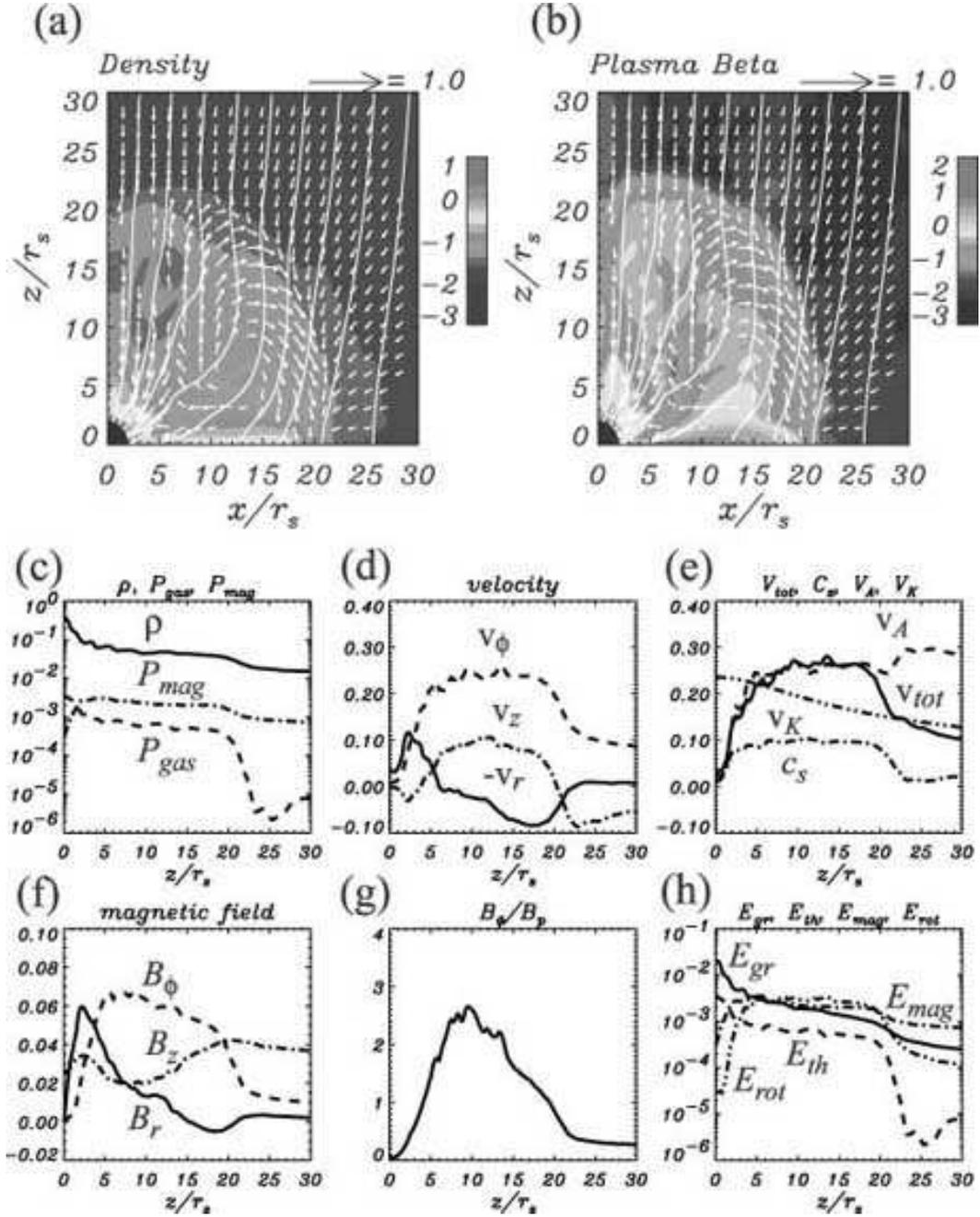}
\caption{Weak magnetic field case A1 ($B_{0} = 0.03$, $v_{0} = 0.01$). 
{\it (a, b)} Snapshots of density and plasma beta at $ t/\tau_{S} = 170$. {\it (c, d, e, f, g, h)} Various physical quantities on the surface of $x/r_{S} = 10$ at $t/\tau_{S} = 170$. The jetlike outflow is stronger (the maximum poloidal velocity of the jet is $\sim 0.1$ c ) than in the standard case A3. The magnetic twist is also larger. \label{fig11}}
\end{figure}

The opposite is true for weaker magnetic fields. Shown in Figure \ref{fig11} is model A1 ($B_{0} = 0.03, v_{0} = 0.01$). The jetlike outflow is clearly stronger than in the standard case A2. The maximum poloidal velocity of the jet becomes about 0.1c. The total velocity exceeds the escape velocity inside the shock. The magnetic twist is significantly stronger, and as a result the propagating shock wave is stronger as well. 

\begin{figure}
\epsscale{0.35}
\plotone{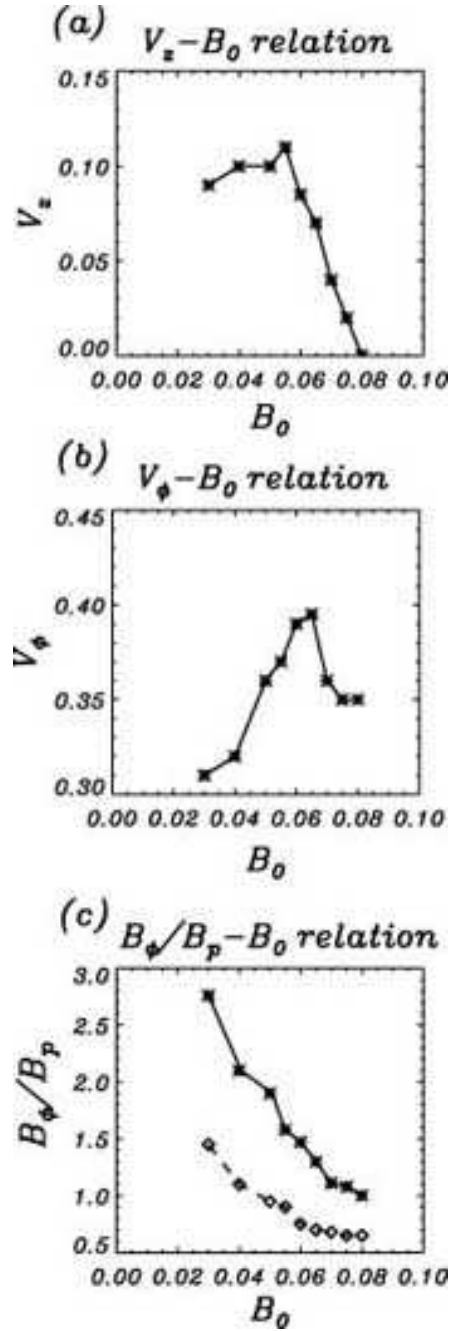}
\caption{Dependence of {\it (a)} the maximum vertical velocity of the jet ($v_{z}$), {\it (b)} the maximum toroidal velocity of the jet ($v_{\phi}$), and {\it (c)} the maximum ratio of toroidal to poloidal magnetic field component ($B_{\phi} / B_{p}$) of the jet {\it (solid line)} and of Alfv\'{e}n waves {\it (dashed line)} on the initial magnetic field strength ($B_{0}$). \label{fig12}}
\end{figure}

Figure \ref{fig12} summarizes these dependenices on the initial magnetic field strength. As the initial magnetic field strength increases, the jet velocity ($v_{z}$, $v_{\phi}$) increases and the magnetic twist decreases. However, for stronger magnetic field ($B_{0} > 0.055$) the jet velocity decreases with increasing $B_{0}$ and the magnetic twist still continues to decrease.  
In order to produce a strong shock, the magnetic field has to be twisted significantly so that it can store enough magnetic energy. If the initial magnetic field is strong, the magnetic field cannot be twisted significantly because Alfv\'{e}n waves propagate as soon as the magnetic field is twisted a little bit. As a result, the jet velocity does not rise up and the magnetic twist remains weak. Therefore, weaker initial magnetic fields are favorable for a stronger jet. 

Let us now consider the physical reason of the above results.
In the Newtonian case, the time evolution of the toroidal magnetic field is given as
\begin{equation}
 {\partial B_{\phi} \over \partial t} = \omega B_{p} ,
\end{equation}
where 
\begin{equation}
\omega \sim {1 \over t_{rot}}
\end{equation}
 and $t_{rot}$ is the rotation time scale.
From this we obtain
\begin{equation}
{B_{\phi} \over B_{p}} \sim \omega t. \label{eq12}
\end{equation}
In this situation, the comparing time scale is the propagation time scale of the Alfv\'{e}n wave 
\begin{equation}
t \sim {L \over v_{Az}} \sim {L \sqrt{4 \pi \rho} \over B_{z}}, \label{eq13}
\end{equation}
where $L$ is the typical length and $v_{Az}$ is the vertical component of the Alfv\'{e}n wave velocity vector.
Substituting equation (\ref{eq13}) for equation (\ref{eq12}), we have
\begin{equation}
{B_{\phi} \over B_{p}} \sim {\omega L \sqrt{4 \pi \rho} \over B_{z}} \propto {1 \over B_{z}}. 
\end{equation} 
In this simulation, $B_{z} \sim B_{0}$. Therefore, 
\begin{equation}
{B_{\phi} \over B_{p}} \sim {1 \over B_{0}} .
\end{equation}
This can explain the dependence of the magnetic twist on the initial magnetic field strength, especially in the case of the Alfv\'{e}n wave component (see Fig. \ref{fig11}c, {\it dashed line}).

The upward motion of the fluid behind the twist wave front is induced by the $\mathbf{J} \times \mathbf{B}$ force. If we neglect other forces, the equation of motion for the fluid element in the z-direction becomes
\begin{equation}
\rho {\partial v_{z} \over \partial t} \sim \nabla \left( {B_{\phi}^{2} \over 8 \pi} \right),
\end{equation}
which can be rewritten as
\begin{equation}
v_{z} \sim {1 \over \rho}{t \over z} \left( {B_{\phi}^{2} \over 8 \pi} \right).\label{eq14}
\end{equation}
In this situation the time scale is determined by the propagation time scale of the Alfv\'{e}n wave. Thus, $t / z \sim 1/v_{Az}$. Since in the region near the wave front, $B_{p} \sim B_{z} \sim B_{0}$, 
equation (\ref{eq14}) can be rewritten as
\begin{equation}
v_{z} \sim {1 \over \rho} {1 \over v_{Az}} \left( {B_{\phi}^{2} \over 8 \pi} \right) \propto B_{0}^{-1}.
\end{equation}
This explains the dependence of $v_{z}$ on $B_{0}$ for $B_{0} > 0.055 $ in Figure \ref{fig11}a; i.e., as the initial magnetic field strength increases, the vertical component of the jet velocity decreases. However, the jet velocity has a limit when the initial magnetic field strength becomes even weaker. It can be understood as follows: the energy transport by the magnetic field is the Poynting flux\begin{equation}
{c \over 4 \pi} {\mathbf{E} \times \mathbf{B}} \sim {B_{\phi}^{2} \over 4 \pi} v_{z} + {B_{z} B_{\phi} \over 4 \pi} v_{\phi}. \label{eq15}
\end{equation}
The first term on the right-hand side of equation (\ref{eq15}) is the jet component, whereas the second term is the Alfv\'{e}n wave component. 
This equation also can explain the dependence of the jet velocity on the initial magnetic field strength. When the initial magnetic field strength becomes stronger,  $B_{z}$ increases. This means that the Alfv\'{e}n wave component takes away more energy than the jet component. Thus, a less magnetic energy is converted to the kinetic energy of the jet, so that the jet becomes weaker.
When the initial magnetic field strength becomes weaker, the magnetic twist becomes stronger and $B_{\phi}$ increases. Thus, much of the magnetic energy is converted to the kinetic energy of the jet. However, the converted magnetic energy has a limit. When the magnetic twist ($B_{p}/B_{\phi}$) of the Alfv\'{e}n wave becomes 1, it becomes a limit. This threshold is determined by the balance between the jet component and the Alfv\'{e}n wave component
\begin{equation}
{B_{\phi}^{2} \over 4 \pi}v_{z} = {B_{z} B_{\phi} \over 4 \pi} v_{\phi},
\end{equation}
which can be rewritten as
\begin{equation}
{B_{\phi} \over B_{z}} = {v_{\phi} \over v_{z}} \sim 1.
\end{equation}
Thus, we can understand that when the magnetic twist in the Alfv\'{e}n wave becomes 1, the jet velocity becomes maximum. 
 
On the other hand, when the initial magnetic field is zero (case D), the shock wave is not generated and the jetlike outflow is not formed either. This clearly demonstrates that the generation of the shock wave, as well as the jetlike outflow, requires magnetic fields. 

\subsection{Dependence on the Initial Rotational Velocity}

\begin{figure}
\epsscale{0.85}
\plotone{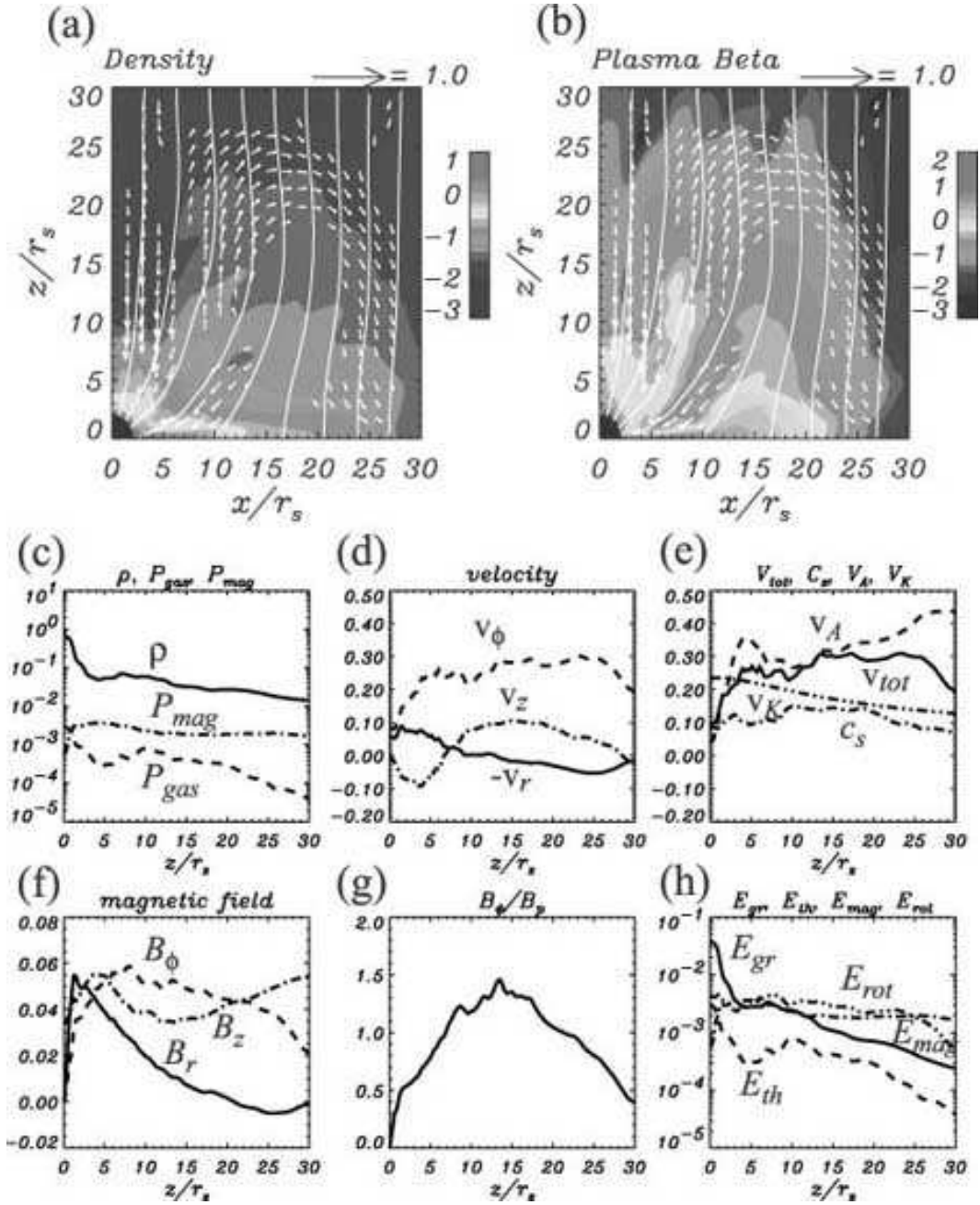}
\caption{Results of the fast rotation case B2 ($B_{0} = 0.05$, $v_{0} = 0.015$). {\it (a, b)} Density and plasma beta ($= P_{gas}/P_{mag} $) at $t/\tau_{S} = 170$. {\it (c, d, e, f, g, h)} Various physical quantities on the surface of $x/r_{S} = 10$ at $t/\tau_{S} = 170$. The jetlike outflow is stronger and wider (the maximum poloidal velocity of the jet is $\sim 0.1$ c ) than in the standard case.  \label{fig13}}
\end{figure}

Next we show the dependence of the jet properties on the initial rotational velocity.
Figure \ref{fig13} shows the results (density and plasma beta distributions) of the fast rotation case B2 ($B_{0} = 0.05, v_{0} = 0.015$). The jetlike outflow is stronger and wider than in the standard case A2. The maximum poloidal velocity of the jet is about 0.1 c. The jet velocity exceeds the local escape velocity in the whole region. The magnetic twist is stronger accordingly. 

\begin{figure}
\epsscale{0.85}
\plotone{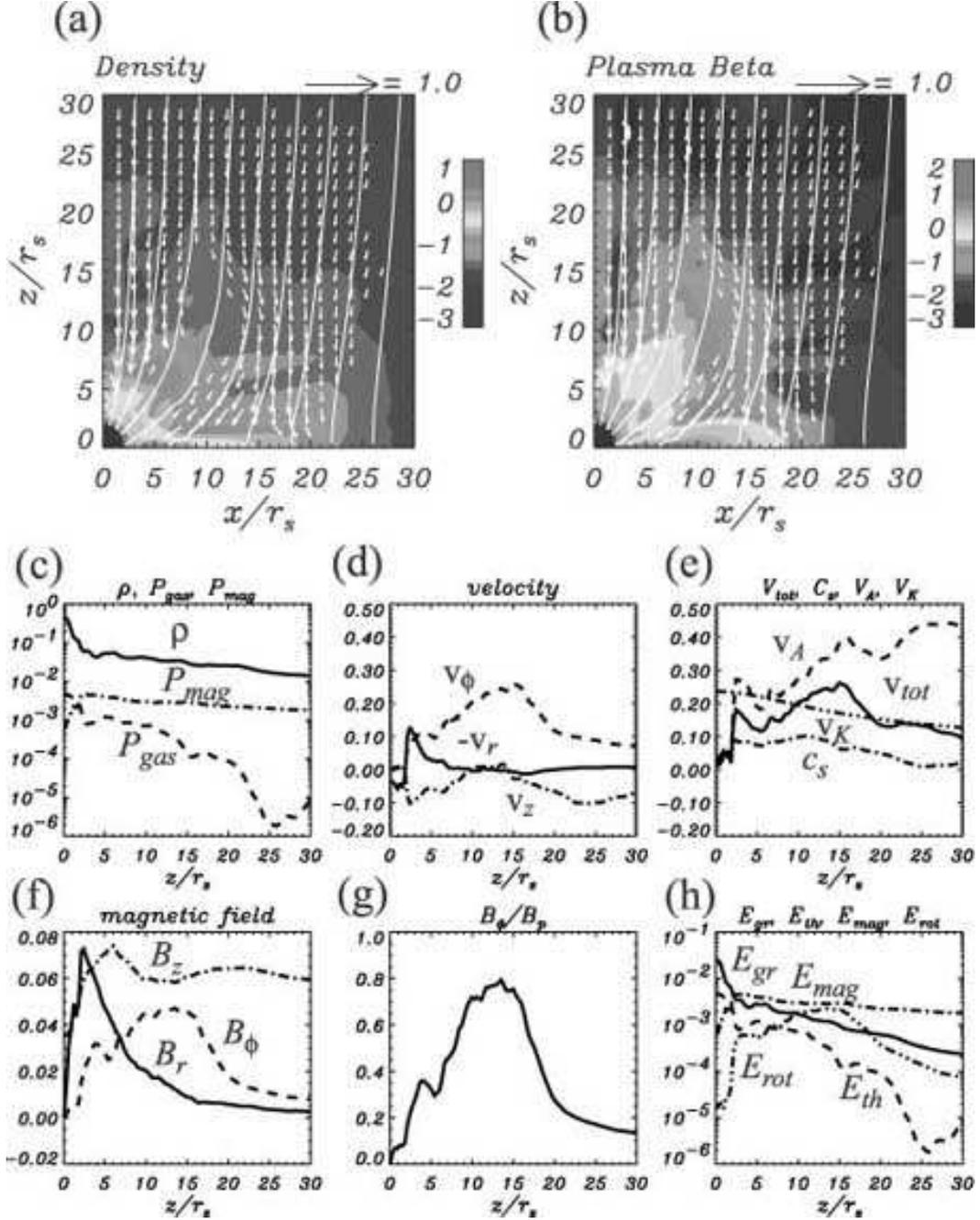}
\caption{Results of the slow rotation case B1 ($B_{0} = 0.05$, $v_{0} = 0.005$). {\it (a, b)} Density and plasma beta at $t/\tau_{S} = 150$. {\it (c, d, e, f, g, h)} Various physical quantities on the surface of $x/r_{S} = 10$ at $t/\tau_{S} = 150$. A jetlike outflow is not generated.  \label{fig14}}
\end{figure}

The slow rotation case B1 ($B_{0} = 0.05, v_{0}=0.005$) is shown in Figure \ref{fig14}. The jetlike outflow is weaker and fainter. The maximum poloidal velocity of the jet is less than 0.01 c. The jet velocity is barely larger than the escape velocity in some regions. The magnetic twist is weaker correspondingly. 

\begin{figure}
\epsscale{0.35}
\plotone{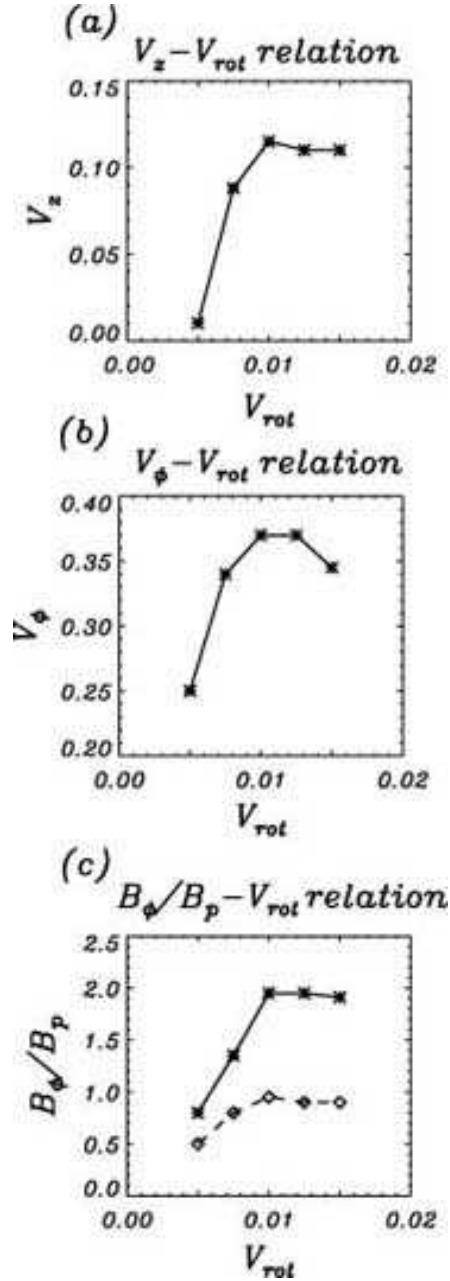}
\caption{Dependence of {\it (a)} the maximum vertical velocity of the jet ($v_{z}$), {\it (b)} the maximum toroidal velocity of the jet ($v_{\phi}$), and {\it (c)} the maximum ratio of toroidal to poloidal magnetic field component ($B_{\phi}/ B_{p}$) of the jet {\it (solid line)} and of Alfv\'{e}n wave {\it (dashed line)} on the initial rotational velocity ($v_{rot}$). \label{fig15}}
\end{figure}

Figure \ref{fig15} shows the dependence of various physical quantities of the jet on the initial rotational velocity.
As the initial rotational velocity becomes faster, the jet velocity ($v_{z}$, $v_{\phi}$) becomes faster and the magnetic twist becomes stronger up to a certain value ($v_{0} = 0.01$). For faster initial rotational velocity ($v_{0} > 0.01$), the vertical component of the jet velocity ($v_{z}$) and the magnetic twist are almost constant and the toroidal component of the jet velocity ($v_{\phi}$) becomes a little bit slower. If the initial rotation is sufficiently fast, the magnetic field is twisted significantly and stores enough energy to produce a fast jet. However, the stored magnetic energy has a limit. It is determined by the competition between the propagation time of Alfv\'{e}n waves and the rotation time of the disk (or the twisting time). This can be understood by the same reason as the dependence on the initial magnetic field strength. The limit is determined when the magnetic twist of the Alfv\'{e}n wave becomes 1. 
 Provided with the results of this and the previous sections, the magnetic field strength is the dominant factor to determine the maximum jet velocity. It should be noted, however, that when the initial rotational velocity is zero (case C), the shock wave is not generated and the jetlike outflow is not formed either. Hence, the magnetic twist is crucial to produce the shock wave and the jetlike outflow.   

\section{Summary and Discussion}

We have studied the generation of a jet from gravitational collapse of a rotating star with magnetic fields by using the 2.5-dimensional GRMHD simulation code. Our results are summarized as follows:

\begin{enumerate}
	\item When the stellar matter falls onto the black hole, the collapse is anisotropic because of the effect of rotation and magnetic field and a disklike structure is formed near the black hole. A  shock wave is launched near the black hole and propagates outward. The jetlike outflow is produced inside the shock. The maximum velocity of the jet is mildly relativistic ($\sim 0.3$ c). This result is consistent with the pseudo-Newtonian case (Proga et al. 2003). The kinetic energy of the jet is $\sim 10^{49} \ \mathrm{ergs} $. 
	
	\item The shock wave is generated mainly by magnetic and centrifugal forces. The frozen magnetic field is twisted and amplified by the rotation of stellar matter particularly near the black hole. It starts to expand outwardly with Alfv\'{e}n waves and leads to the production of the shock wave. 
	
	\item The acceleration of the jet is also driven by the magnetic pressure and centrifugal forces. The jet is collimated in the course of propagation by the pinching force of the toroidal magnetic field and the geometry of the poloidal magnetic field. 
	
	\item The magnetic twist is large in the jet-forming region. The jet has a helical structure. This is similar to other astrophysical jets formed by MHD processes. 
	
	\item The Poynting flux is an order of magnitude larger than the kinetic energy flux. Alfv\'{e}n waves as the Poynting flux transports more energy outward than the jet.
	
	\item As the initial magnetic field strength increases, the jet velocity increases and the magnetic twist decreases. However, for stronger magnetic field the jet velocity decreases with increasing initial magnetic field strength and the magnetic twist goes on decreasing. The dependence of the jet properties on the initial rotational velocity is similar to the dependence on the initial magnetic field strength. As the initial rotational velocity becomes faster, the jet velocity becomes faster and the magnetic twist becomes stronger up to a certain value. For faster initial rotational velocity, the jet velocity and the magnetic twist are almost constant. The magnetic energy is converted to the kinetic energy of the jet. However, the converted magnetic energy has a limit. It is determined by the competition between the propagation time of Alfv\'{e}n waves and the rotation time of the disk (or the twisting time). Hence the jet velocity is saturated at some point.
\end{enumerate}

The maximum jet velocity in the current simulations was about $0.3$ c and the typical kinetic energy of the jet was $\sim 10^{49} \  \mathrm{ergs}$ . 
The jet is too slow and too weak for the jet of GRBs. However, the baryon mass in the jet is small. This gives good results for the GRB model. 
We have to consider other acceleration mechanisms.
We expect that disk jets formed from an accretion disk become faster than the jet in the current simulations because they can release more gravitational energy and have enough energy to be applied to the GRB model. We believe that, if the current simulations can be continued for longer times, a disk-jet may be formed by the MHD processes.
If the jet contains enough energy to convert the kinetic energy of the jet, then the break-out of the jet through the stellar surface is a possibility. 
When the jet goes through the stellar surface, the strong density gradient may accelerate the jet. In fact, some authors (Aloy et al. 2000; Zhang, Woosley, \& MacFadyen 2003) have shown numerically that a significant acceleration of the jet occurs and the terminal Lorentz factor becomes as high as $\Gamma \sim 50$.
 Since they used special relativistic hydrodynamics codes and neglected magnetic fields altogether, we think that it is important to simulate the propagation of the jet outside the stellar surface by a GRMHD code properly evaluating the importance of magnetic fields for the dynamics and further propagation of the jet.

On the other hand, our results may apply to baryon-rich outflows associated with the so-called failed GRBs. It is supposed to be a high baryon-load fireball attaining mildly relativistic velocities and not producing GRBs. Such failed GRBs may have event rates higher than successful GRBs (Woosley et al. 2002; Huang, Dai, \& Lu 2002). Some failed GRBs can be observed as \lq\lq hypernovae." SN 2002ap may be such an event. It shows broad-line spectral features like the famous SN 1998bw (Kinugasa et al. 2002), yet it is not associated with a GRB. Kawabata et al. (2002) proposed a jet model for SN 2002ap consistent with their spectropolarimetric observations. Totani (2003) estimated that the jet should be moving at about 0.23 c in the direction perpendicular to the line of sight and that the kinetic energy of the jet was $\sim 5 \times 10^{50}\  \mathrm{ergs}$ .

We have assumed a uniform global magnetic field in the simulations. However, this assumption may be unrealistic. From observations, it is inferred that rotating compact stars in general have a dipole-like magnetic field. Hence, it may be more likely that the rotating stars collapse with a dipole-like magnetic field. On the other hand, it is possible that the magnetic field near the central black hole is radial because of radial accretion of matter. 
We will study these possibilities in a forthcoming paper.
There are a lot of works on astrophysical jets using non-relativistic MHD simulations with a dipole magnetic field configuration (e.g., Hayashi, Shibata, \& Matumoto 1996; Goodson, Winglee, \& B\"{o}hm  1997; Goodson, B\"{o}hm, \& Winglee 199; Goodson \& Winglee 1999; Romanova et al. 2002; Kato et al. 2003). These simulations show a high-velocity wind and lead the ejection of hot plasmoids. If the plasmoids are ejected intermittently, they will collide with each other, forming \lq\lq internal shocks," which may explain the time variability of GRBs (Shibata \& Aoki 2003; Aoki, Yashiro, \& Shibata. 2003).

\medskip

Y. M. appreciates many helpful conversations on GRBs with S. Aoki, M. Uemura, and T. Totani. 
He also thanks K. Watarai, Y. Kato, K. Uehara, K. Nishikawa, and T. Haugboulle for useful discussions. This work was partially supported by Japan Science and Technology Cooperation (ACT-JST), Grant-in-Aid for the 21st Century COE ``Center for Diversity and Universality in Physics'' and Grants-in-Aid for the scientific
research from the Ministry of Education, Science, Sports, Technology, and Culture of Japan through 14079202, 14540226 (PI: K. Shibata) and 14740166.
The numerical computations were partly carried out on the VPP5000 at the Astronomical Data Analysis Center of the National Astronomical Observatory, Japan (yym17b), 
and partly on the Alpha Server ES40 at the Yukawa Institute for Theoretical Physics of Kyoto University, Japan.

\begin{deluxetable}{llc}
\tablecolumns{3}
\tablewidth{0pc}
\tablecaption{Units and Normalization \label{table1}}
\tablehead{
\colhead{Phisical Quantity} & \colhead{Description} & \colhead{Normalization Unit}}
\startdata
$\tau_{S}$ & Time       & $r_{S}/c$ \\
$r,z$        & Length         & $r_{S}$ \\
$\rho$     & Density        & $\rho_{0}$ \\
$P$        & Pressure       & $\rho_{0} c^{2}$ \\
$v$        & Velocity       & $c$ \\
$B$        & Magnetic field & $\sqrt{\rho_{0} c^{2}}$ 
\enddata
\tablecomments{The unit of length is the Schwarzschild radius $r_{S} \equiv 2GM_{\odot}/c^{2}$. The unit of density $\rho_{0}$ is the initial density at $(r , z) = (3 r_{S} , 0)$. }
\end{deluxetable}

\begin{deluxetable}{lccccc}
\tablecolumns{6}
\tablewidth{0pc}
\tablecaption{Models and Parameters \label{table2}}
\tablehead{
\colhead{Case} & \colhead{$B_{0}$} & \colhead{$v_{0}$} & \colhead{$E_{th}$} & \colhead{$E_{mag}$} & \colhead{$E_{rot}$}} 
\startdata
A1 &  0.03      &  0.01 & $2.36 \times 10^{-3}$ & $6.04 \times 10^{-4}$ & $5.36 \times 10^{-2}$\\
A2 &  0.05      &  0.01 & $2.36 \times 10^{-3}$ & $1.68 \times 10^{-3}$ & $5.36 \times 10^{-2}$\\
A3 &  0.07      &  0.01 & $2.36 \times 10^{-3}$ & $3.28 \times 10^{-3}$ & $5.36 \times 10^{-2}$\\
B1 &  0.05      &  0.005 & $2.36 \times 10^{-3}$ & $1.68 \times 10^{-3}$ & $2.68 \times 10^{-2}$\\
B2 &  0.05      &  0.015 & $2.36 \times 10^{-3}$ & $1.68 \times 10^{-3}$ & $8.03 \times 10^{-2}$\\
C  &  0.05      &  0.0   & $2.36 \times 10^{-3}$ & $1.68 \times 10^{-3}$ & 0.00 \\
D  &  0.0       &  0.01  & $2.36 \times 10^{-3}$ & 0.00 & $5.36 \times 10^{-2}$
\enddata
\tablecomments{Case A2 ($B_{0} = 0.05$, $v_{0}=0.01$) is considered to be the standard case in our simulations. Cases \lq\lq A" differ from case A2 in initial magnetic field strength and cases \lq\lq B" differ from A2 in the initial rotational velocity. In case \lq\lq C" there is no rotation while in case \lq\lq D" there is no magnetic field.}
\end{deluxetable}


\begin{thebibliography}{}

\bibitem[Aloy et al.(2000)]{Alo00} Aloy, M. A., Muller, E., Ibanez, J. M., Marti, J. M., \& MacFadyen, A. I.  2000, \apjl, 531,  L119
\bibitem[Aoki, Yashiro, \& Shibata(2003)]{Aok03} Aoki, S., Yashiro, S, \& Shibata, K.  2003, preprint
\bibitem[Ardeljan, Bisnovatyi-Kogan, \& Moiseenko(2000)]{Ard00} Ardeljan, N. V., Bisnovatyi-Kogan, G. S., \& Moiseenko, S. G.   2000, \apss, 274, 389
\bibitem[Bloom, Kulkarni, \& Djprgovski(2002)]{Blo02a} Bloom, J. S., Kulkarni, S. R., \& Djorgovski G.  2002, \aj, 123, 1111
\bibitem[Bloom et al(2002)]{Blo02b} Bloom, J. S., et al. 2002, \apjl, 572, L45
\bibitem[Bloom, Frail, \& Kulkarni(2003)]{Blo03} Bloom, J. S., Frail, D. A., \& Kulkarni, S. R.  2003, \apj, 594, 674
\bibitem[Bruenn(1992)]{Bru(92)} Bruenn, S. W.  1992, in Neuclear Physics in the Universe, ed. M. W. Guidry \& M. R. Strayer (Philadelphia: Institute of Physics Pub.), 31 
\bibitem[Cameron(2001)]{Cam01} Cameron, A. G. W.   2001, \apj, 562, 456
\bibitem[Davis(1984)]{Dav84} Davis, S. F.   1984, NASA Contractor Rep. 172373, ICASE Rep., No. 84-20
\bibitem[De Villiers \& Hawley(2003)]{DeV03} De Villiers, J. -P. \& Hawley, J. F. 2003, \apj, 589, 458
\bibitem[Drenkhahn \& Spruit(2002)]{Dre02} Drenkhahn, G. \& Spruit, H. C.  2002,  \aap, 391, 1141
\bibitem[Fishman \& Meegan(1995)]{Fis95} Fishman, G. J. \& Meegan, C. A.  1995, \araa, 33, 415
\bibitem[Frail(2001)]{Fra01} Frail, D., et al. 2001, \apjl, 562, L55
\bibitem[Gammie, McKinney, \& T\'{o}th(2003)]{Gam03} Gammie, C. F., McKinney, J. C., \& T\'{o}th, G. 2003, \apj, 589, 444
\bibitem[Garnavich(2003)]{Gar03} Garnavich, P. M., et al. 2003, \apj, 582, 924
\bibitem[Goodson, Winglee, \& B\"{o}hm(1997)]{God97} Goodson, A. P., Winglee, R. M., \& B\"{o}hm, K. H.   1997, \apj, 489, 199
\bibitem[Goodson, B\"{o}hm, \& Winglee(1999)]{God99a} Goodson, A. P., B\"{o}hm, K. H., \& Winglee, R. M.   1999, \apj, 524, 142
\bibitem[Goodson \& Winglee(1999)]{God99b} Goodson, A. P. \& Winglee, R. M.   1999, \apj, 524, 159
\bibitem[Hawley, Smarr, \& Wilson(1984)]{Haw03} Hawley, J. F., Smarr, L. L., \& Wilson, J. R. 1984, \apjs, 55, 211
\bibitem[Hayashi, Shibata, Matumoto(1996)]{Hay96} Hayashi, M. R., Shibata, K., \& Matumoto, R.   1996, \apjl, 468, L37
\bibitem[Heger et al.(2003)]{Heg03} Heger, A., Woosley, S. E., Langer, N, \& Spruit, H. C. 2003, preprint (astro-ph/0301374)
\bibitem[Hjorth et al.(2003)]{Hjo03} Hjorth, J., et al.  2003, \nat, 423, 847
\bibitem[Huang, Dai, \& Lu(2002)]{Hua02} Huang, Y. F., Dai, Z. G., \& Lu, T. 2002, \mnras, 332, 735
\bibitem[Iwamonoto et al.(1998)]{Iwa98} Iwamoto, K., et al. 1998, \nat, 395, 672
\bibitem[Kato, Kudoh, \& Shibata(2002)]{Kat02} Kato, S. X., Kudoh, T., \& Shibata, K.  2002, \apj, 565, 1035
\bibitem[Kato, Mineshige, \& Shibata(2003)]{Kat03} Kato. Y., Mineshige, S., \& Shibata, K. 2003, \apj, accepted (astro-ph/0307306)
\bibitem[Katz(1997)]{Kat97} Katz, J. I. 1997, \apj, 490, 633
\bibitem[Kawabata et al.(2002)]{Kaw02} Kawabata, K. S., et al.  2002, \apjl, 580, L39
\bibitem[Kinugasa et al.(2002)]{Kin02} Kinugasa, K., et al.  2002, \apjl, 577, L97
\bibitem[Khokhlov et al.(1999)]{Kho99} Khokhlov, A. M., H\"{o}flich, P. A., Oren, E. S., Wheeler, J. C., Wang, L., \& Chtchelkanova, A. Yu.  1999, \apjl, 524, L107
\bibitem[Klebesadel, Strong, \& Olson(1973)]{Kle73} Klebesadel, R. W., Strong, I. B., \& Olson, R. A.  \apjl,  182,  L85
\bibitem[Klu\'{z}niak \& Ruderman(1998)]{Klu98} Klu\'{z}niak, W. \& Ruderman, M.   1998,  \apjl,  505,  L113
\bibitem[Koide, Shibata, \& Kudoh(1998)]{Koi98} Koide, S., Shibata, K., \& Kudoh, T.  1998, \apjl,  495,  L63
\bibitem[Koide, Shibata, \& Kudoh(1999)]{Koi99} ------.  1999, \apj,  522,  727
\bibitem[Koide (2003)]{Koi03} Koide, S.  2003, \prd,  67, 104010
\bibitem[Kudoh, Matsumoto, \& Shibata(1998)]{Kud98} Kudoh, T., Matsumoto, R., \& Shibata, K. 1998, \apj, 508, 186
\bibitem[LeBlanc \& Wilson(1970)]{Leb70} LeBlanc, J. M. \& Wilson, J. R. 1970, \apj, 161, 541
\bibitem[Lyutikov \& Blackman(2001)]{Lyu01} Lyutikov, M. \& Blackman, E. G.  2001,  \mnras, 321,  177
\bibitem[MacFadyen \& Woosley(1999)]{Mac99} MacFadyen, A. I. \& Woosley, S. E.  1999, \apj, 524, 262
\bibitem[MacFadyen ,Woosley, \& Heger(2001)]{Mac02} MacFadyen, A. I., Woosley, S. E. \& Heger, A.  2001, \apj, 550, 410 
\bibitem[M\'{e}sz\'{a}ros \& Rees(1997)]{Mez97} M\'{e}sz\'{a}ros, P. \& Rees, M. J.  1997, \apjl, 482, L29
\bibitem[M\'{e}sz\'{a}ros(2002)]{Mez02} M\'{e}sz\'{a}ros, P.  2002 \araa, 40, 137
\bibitem[M\"{o}nchmeyer \& M\"{u}ller (1989)]{Mon89} M\"{u}nchmeyer, R. M. \& M\"{u}ller, E. 1989, in NATO ASI Series C262, Timing Neutron Stars, ed. H. Ogelman \& E.P.J. Van den Heuvel (New York: ASI), 549
\bibitem[Panaitescu \& Kumar(2001)]{Pan01} Panaitescu, A. \& Kumar, P. 2001, \apjl, 560, L49
\bibitem[Popham, Woosley, \& Fryer(1999)]{Pho99} Popham, R., Woosley, S. E., \& Fryer, C. 1999 \apj, 518, 356
\bibitem[Piran(1999)]{Pir99} Piran, T. 1999, \physrep, 314, 575
\bibitem[Price et al.(2003)]{Pri03} Price, P. A., et al.  2003,  \nat, 423, 844
\bibitem[Proga et al.(2003)]{Pro03} Proga, D., MacFadyen, A. I., Armitage, P. J., \& Begelman, M. C. 2003, \apjl, 599, L5
\bibitem[Romanova et al.(2002)]{Rom02} Romanova, M. M., Ustyugoya, G. V., Koldoba, A. V., \& Lovelace, R. V. E.  2002, \apj, 578, 420 
\bibitem[Sari, Piran, \& Halpern(1999)]{Sar99} Sari, R., Piran, T., \& Halpern, J. P.  1999, \apjl, 517, L17
\bibitem[Shibata \& Aoki(2003)]{Shi03} Shibata, K. \& Aoki, S.  2003, preprint (astro-ph/0303253)
\bibitem[Shibata \& Uchida(1986)]{Shi86} Shibata, K. \& Uchida, Y.  1986, \pasj, 38, 631
\bibitem[Shibata \& Uchida(1990)]{Shi90} --------. 1990 ,\pasj, 42, 39
\bibitem[Spruit, Daigne, \& Drenkhahn(2001)]{Spu(01)} Spruit, H. C., Daigne, F., \& Drenkhahn, G. 2001, \aap, 369, 694
\bibitem[Stanek et al.(2003)]{Sta03} Stanek, K. Z., et al. 2003, \apjl, 591, L17
\bibitem[Symbalisty(1984)]{Sym84} Symbalisty, E. M. D. 1984, \apj, 285, 729
\bibitem[Thompson(1994)]{Tom94} Thompson, C. 1994, \mnras, 270, 480
\bibitem[Thorne, Price, \& Macdnald(1986)]{Tho86} Thorne, Kip S., Price, R. H., \& Macdonald, D. A.  1986, Membrane Paradigm (Yale University Press, New Haven and London)
\bibitem[Totani(2003)]{Tot03} Totani, T.  2003, \apj, in press (astro-ph/0303621)
\bibitem[Uchida \& Shibata(1985)]{Uch85} Uchida, Y. \& Shibata, K.  1985, \pasj, 37, 515
\bibitem[Uemura et al.(2003)]{Uem03} Uemura, M., et al.  2003, \nat,  423, 843
\bibitem[Usov(1994)]{Uso94} Usov, V. V.  1994, \mnras, 267, 1035
\bibitem[van Paradijs, Koubeliotou, \& Wijers (2000)]{vPa00} van Paradijs, J., Koubelitou, C., \& Wijers, Ralph A. M. J.  2000, \araa, 38, 379
\bibitem[Vlahakis \& K\"{o}nigl(2001)]{Vla01} Vlahakis, N. \& K\"{o}nigl, A.  2001, \apjl, 563, L129
\bibitem[Vlahakis \& K\"{o}nigl(2003a)]{Vla03a} ------. 2003a, \apj, 596, 1080
\bibitem[Vlahakis \& K\"{o}nigl(2003b)]{Vla03b} ------. 2003b, \apj, 596, 1104
\bibitem[Weinberg(1972)]{Wei72} Weinberg, S.  1972, Gravitation and Cosmology (wiely, New York)
\bibitem[Wheeler, Meier, \& Wilson(2002)]{Whe02} Wheeler, J. C., Meier, D. L., \& Wilson, J. R.   2002, \apj, 568, 807
\bibitem[Woosley(1993)]{Woo93} Woosley, S. E.  1993, \apj, 405, 273
\bibitem[Woosley, Eastman, \& Schmidt(1999)]{Woo99} Woosley, S. E., Eastman, R. G., \& Schmidt, B. 1999, \apj, 516, 788
\bibitem[Woosley et al.(2002)]{Woo02} Woosley, S. E., Zhang, W. \& Heger, A. 2002, preprint(astro-ph/0211063)
\bibitem[Yamada \& Sato(1994)]{Yam94} Yamada, S. \& Sato, K.  1994, \apj, 434, 268
\bibitem[Zhang, Woosley, \& MacFadyen(2003)]{Zan03} Zhang, W., Woosley, S. E., \& MacFadyen, A. I.  2003, \apj, 586, 356


\end{thebibliography}
\end{document}